\documentclass[review,12pt,times,authoryear]{elsarticle}
	\usepackage[table]{xcolor}
	\usepackage[colorlinks = true, linkcolor = blue, urlcolor = blue, citecolor = blue, anchorcolor = blue]{hyperref}
	\usepackage{diagbox}
	\usepackage{multirow}
	\usepackage{fancyhdr}
	\usepackage{amsmath}
    \usepackage{amssymb}
	\usepackage{indentfirst}
	\usepackage{booktabs}
	\usepackage{graphicx}
	\usepackage{float}
	\usepackage{subfigure}
	\usepackage[section]{placeins}
    \usepackage{geometry}
    \usepackage{flushend,cuted}
    \usepackage{bm}
    \usepackage{caption}
\usepackage{xcolor,soul}
\usepackage{array}
\usepackage{url}
\newcolumntype{P}[1]{>{\centering\arraybackslash}p{#1}}
\definecolor{darkgreen}{rgb}{0.0,0.5,0.0}
\definecolor{lavender}{rgb}{0.5,0.25,0.5}
\definecolor{darkred}{rgb}{0.85,0.0,0.15}
\makeatletter
\newcommand{\tensor}[1]{\mathbf{#1}}

\newcommand{\Rmnum}[1]{\expandafter\@slowromancap\romannumeral #1@}
\makeatother

\newcommand{\revision}[1]{\textcolor{black}{#1}}

\journal{arXiv}

\begin{document}

\begin{frontmatter}

\title{Prediction of Yield Surface of Single Crystal Copper from Discrete Dislocation Dynamics and Geometric Learning}

\author[1]{Wu-Rong Jian}

\author[2]{Mian Xiao}

\author[2]{WaiChing Sun\corref{corr-author}}
\ead{wsun@columbia.edu}

\author[1]{Wei Cai\corref{corr-author}}
\ead{caiwei@stanford.edu}

\address[1]{Department of Mechanical Engineering, Stanford University, Stanford CA, 94305, USA}

\address[2]{Department of Civil Engineering and Engineering Mechanics, Columbia University, 614 SW Mudd, 4709, New York, NY 10027, USA}

\cortext[corr-author]{Corresponding author}

\date{\today}

\begin{abstract}
The yield surface of a material is a criterion at which macroscopic plastic deformation begins. For crystalline solids, plastic deformation \revision{occurs through} the motion of dislocations, which can be captured by discrete dislocation dynamics (DDD) simulations. In this paper, we predict the yield surfaces and strain-hardening behaviors using DDD simulations and a geometric manifold learning approach. The yield surfaces in the three-dimensional space of plane stress are constructed for single-crystal copper subjected to uniaxial loading along the $[100]$ and $[110]$ directions, respectively.  With increasing plastic deformation under $[100]$ loading, the yield surface expands nearly uniformly in all directions, corresponding to isotropic hardening.
In contrast, under $[110]$ loading, latent hardening is observed, where the yield surface remains nearly unchanged in the orientations in the vicinity of the loading direction itself but expands in other directions, resulting in an asymmetric shape.
This difference in hardening behaviors is attributed to the different dislocation multiplication behaviors on various slip systems under the two loading conditions.
\end{abstract}


\end{frontmatter}

\section{Introduction}
\label{sec:introduction}
Understanding crystal plasticity in terms of fundamental physics has been a long-standing goal in computational materials science. Since the proposal of crystal dislocations~\citep{taylor1934prsla,polanyi1934zp,orowan1934zp} and their observation by transmission electron microscopy~\citep{hirsch1956pm}, it has been well-established that the plastic deformation behaviors of crystalline materials are controlled by the motion of dislocations \citep{hirth1982theory}. The discrete dislocation dynamics (DDD) simulation method \citep{amodeo1990prb,devincre1997msea,arsenlis2007msmse} has been developed to establish the connection between the microscopic motion of individual dislocations and the macroscopic stress-strain behavior of the single-crystalline material.
A fundamental concept in describing the plastic deformation behavior of materials is the yield surface~\citep{hill1948prsla, meyers2008mechanical}. When the local stress at a material point is within the yield surface, the deformation is purely elastic. By contrast, when the local stress reaches the yield surface, plastic deformation begins. Yield surfaces may also evolve due to the evolution of microstructures. Common theories for plasticity include strain- and work-hardening of which the yield surface may evolve due to the accumulation of plastic deformation as well as plastic work. 
For example, isotropic hardening corresponds to the uniform expansion of the yield surface in all directions, while kinematic hardening corresponds to the translation of the yield surface in the stress space without changing its size and shape \citep{hill1998mathematical}.
For metals, the Bauschinger effect \citep{bauschinger1886mmtlm}, a phenomenon \revision{in} which the plastic deformation \revision{accumulating} in one direction may cause a lower threshold along the opposite direction, is commonly idealized via a pure kinematic hardening model while the true plastic behaviors can be geometrically more complex in the stress space. 
Given the importance of yield surface in solid mechanics and the fact that the DDD simulations are performed at the mesoscale suitable for upscaling to continuum scale models, leveraging DDD simulations to deduce yield surface in stress space seems like a logical step toward more precise and accurate constitutive models for \revision{single crystal} plasticity. However, to our best knowledge, there have not been predictions of yield surface from DDD simulations to date. This lack of attempts/successes could be attributed \revision{to} the following technical barriers. 
First, DDD simulations are computationally very expensive. To predict the plastic deformation behavior of a crystal, the motion of dislocations in a large enough simulation cell needs to be followed for a long enough time to accumulate a significant level of plastic strain. In addition, interactions between/among nearby dislocations require a very small time step to maintain numerical stability. 
Second, constructing the yield surface for a given computational sample (i.e., a dislocation configuration) by brute-force DDD simulations would require a very large number of DDD simulations, each with a different loading (stress) orientation. Given that the stress conditions span a six-dimensional (6D) parameter space for a given internal state, the essential number of DDD simulations would be overwhelming if the DDD stress paths were uniformly sampled. Presumably, material symmetry may reduce the dimensionality of the yield surface. However,  the anisotropic nature of slip systems makes it infeasible to project the yield surface onto the principal stress space. 
Thirdly, we need a parameterization to express the yield surface with sufficient expressivity (cf. \citep{raghu2017expressive}) to interpolate data from DDD predictions while generalizing well in the extrapolation regime without overfitting.  
Finally, given the high cost of obtaining data from DDD simulations, a framework that may consistently help deduce stress paths that provide the most valuable information is needed, such that the fidelity of the models may gradually improve with additional DDD simulations.

In this work, we construct the yield surfaces of single-crystal Cu from DDD simulations with the following approaches to overcome the above challenges.
First, recent progress on subcycling integrator \citep{sills2016msmse} and GPU implementation \citep{bertin2019msmse} has made DDD simulations much more efficient than before. For single crystal Cu, the plastic strain on the order of $1\%$ strain can be reached using a single GPU over the time period of a week, although the strain rate still needs to be kept high (e.g., $10^3$\,s$^{-1}$) in these simulations. In this work, we will keep our effective strain rate about $10^3$\,s$^{-1}$.  Hence, our predicted yield surface is not the same as the one typically might have in mind, i.e., for quasi-static loading conditions.  Our work represents a first attempt at this problem, and the strain rate effect on the yield surface will be examined in the future. In addition, we show that if we need to determine the yield point for a given loading direction, the DDD simulation can be much shorter than those needed for determining the strain hardening rate.
Second, to reduce the total number of DDD simulations, we will limit our scope to the \revision{plane stress} condition, \revision{with} only three non-zero stress components. This means we will construct a three-dimensional (3D) ``cross-section'' of the full yield surface, which lives in the 6D space. The yield surface in the 3D plane-stress conditions is also much easier to visualize, and we show that quite interesting insights can be gained by examining how it evolves while the crystal is plastically deformed along different loading directions.
Third, we use the geometric prior method \citep{xiao2022cmame} to construct the yield surface (as a hypersurface) from the DDD simulation data, where each patch with a local coordinate chart provides  a local representation of the yield surface. Together, these patches overlap consistently and offer a general description of any surface via an atlas. 
The geometric prior framework can provide a convenient means to evaluate local features on the yield surface, e.g., local curvature, which can be used to decide where more data should be collected from additional DDD simulations.
Prior to the geometric prior method, the non-uniform rational B-spline (NURBS) method \citep{coombs2016cmame,coombs2017cmame,coombs2018cmame}, the \revision{level set} models \citep{vlassis2021cmame, vlassis2022jam} and the deep neural network based surrogate model \citep{nascimento2023ijp} had also been used to \revision{construct} yield \revision{surfaces}. While these methods have shown great promise in representing the yield surface of complex geometries implicitly, they are not easily updated as the yield function must be defined in the entire stress space. By contrast, a geometric prior represented by the collection of coordinate charts allows models to be updated locally when new data are available. The locality is advantageous for the DDD yield surface calculations where an active learning setting may leverage the update inductive bias to improve fidelity while the yield surface is updated at multiple stages, without re-training the entire model from scratch.

The rest of this paper is organized as follows. First, the DDD simulation setup and the procedure of yield stress data extraction from DDD simulation are described in Section \ref{sec:2.1}, followed by the details of our yield surface construction framework using geometric prior in Section \ref{sec:2.2}. We present the 3D yield surfaces of various dislocation configurations of single-crystal Cu upon plane stress loading in Section \ref{sec:3.1} and the two-dimensional (2D) yield loci for uniaxial loading in Section \ref{sec:3.2}. The results show that the yield surface evolves in different modes during the strain hardening of single-crystal Cu along [100] crystal orientation (isotropic hardening) and [110] crystal orientation (latent hardening), respectively. Section~\ref{sec:4.1} gives an explanation of this observation based on the different dislocation multiplication behaviors on various slip systems for these two loading orientations. \revision{Section~\ref{sec:4.2} discusses the effect of microstructure initialization on yield surfaces.  Section~\ref{sec:4.3}  compares the yield surfaces obtained from DDD simulations with those calculated from the classical yield criteria (von Mises and Tresca).} The outlooks for DDD simulations and the geometric prior method for the yield surface construction of single crystals are provided in Section~\ref{sec:4.4} and Section~\ref{sec:4.5}, respectively. 
The conclusions are given in Section~\ref{sec:conclude}.

\section{Methodology}
\label{sec:methodology}
\subsection{Data collection from DDD}
\label{sec:2.1}
The DDD simulations are conducted using the open-source ParaDiS program \citep{arsenlis2007msmse,ParaDiS}, which utilizes the recently developed sub-cycling time integration scheme \citep{sills2016msmse} and its Graphics Processing Units (GPU) implementation \citep{bertin2019msmse}.
The material parameters in our DDD simulations correspond to single crystal copper, e.g., Burgers vector magnitude $b = 0.255$ nm, Poisson’s ratio $\nu = 0.324$, and shear modulus $\mu = 54.6$ GPa. During all DDD simulations, a linear mobility law with drag coefficient $B = 1.56 \times 10^{-5} \mathrm{~Pa} \cdot \mathrm{s}$ is applied to glissile dislocations. Cross-slip is not allowed, and dislocation junctions are only allowed to move along their line directions using zipping or unzipping mode. All the DDD simulation parameters are also summarized in \autoref{tbl:simParam}. Due to the large amount of simulations necessary to obtain a yield surface in the general 6D stress space, here we limited our attention to plane stress conditions, where only $\sigma_{xx}$, $\sigma_{yy}$, $\sigma_{xy}$ may be non-zero.

\begin{figure}[ht]
    \centering
    \includegraphics[width=0.6\textwidth]{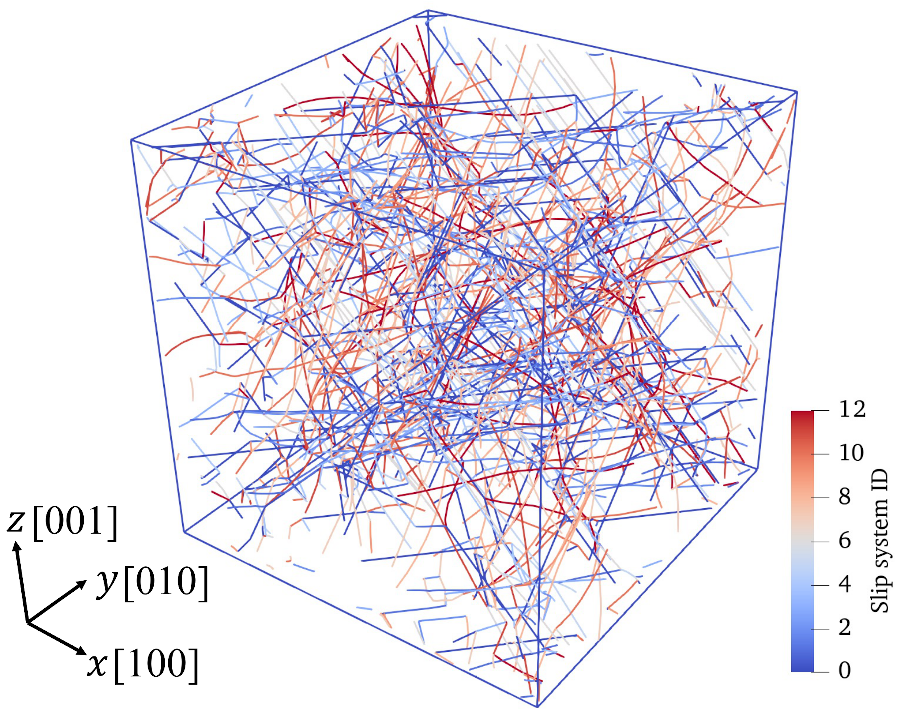}
    \caption{The initial dislocation configuration (\revision{Config}-0) with a dislocation density of $\rho_0 \approx 1.2 \times 10^{12} \mathrm{~m}^{-2}$ before loading in DDD simulation. \revision{The $x$, $y$, $z$-axes are along the $[100]$, $[010]$, $[001]$ crystal orientations, respectively.}}
    \label{fig:initial_config}
\end{figure}

The dimensions for the simulation model are $\sim$ 15 $\mu$m $\times$ 15 $\mu$m $\times$ 15 $\mu$m. To build the initial dislocation configuration for the subsequent loading simulations, straight dislocation lines on the $\frac{1}{2}\langle 110\rangle\{111\}$ slip systems are introduced randomly to the simulation box, where periodic boundary conditions (PBCs) are applied along all three directions. The dislocation configuration is then relaxed, resulting in a dislocation density of $\rho_0 \approx 1.2 \times 10^{12} \mathrm{~m}^{-2}$. The relaxed configuration, shown in \autoref{fig:initial_config} and also used in our previous works \citep{akhondzadeh2020jmps,akhondzadeh2021mt}, is labeled \revision{Config-0}, and considered as the reference configuration in this work. 
%
\revision{The coordinate system in the physical space is selected such that the $x$, $y$, $z$ directions correspond to the $[100]$, $[010]$, $[001]$ crystal orientations of single crystal Cu.}
%

To construct the yield surface of a given (relaxed) dislocation configuration, we perform a set of DDD simulations, each \revision{corresponding to a point on the yield surface in the stress space.}
\revision{While the principal stresses are often used to specify the stress state in classical yield criteria for isotropic material, they are not very useful for describing the yield surfaces of single crystals which are plastically anisotropic.}
\revision{This is because for plasticity in single crystals, not only the principal stress values but also their orientations relative to the crystal axes are important. Specifying the principal stress values and their orientations would require six degrees of freedom, which equals the number of degrees of freedom when directly using the stress components.}
%
%
\revision{Therefore, for single crystals it is more convenient to directly work with the components of the stress in a given coordinate system, such as the one chosen here, where the $x$-$y$-$z$ axes align with the cubic axes of the crystal. This approach has been used in previous works, e.g.~\citep{mecking1996acta}.} 

For simplicity, we focus on the plane stress condition (in the $x$-$y$ plane, i.e., the $(001)$ crystallographic plane) where the non-zero stress components are limited to $\sigma_{xx}$, $\sigma_{yy}$ and $\sigma_{xy}$.
In this case, the magnitude and orientation of the stress tensor $\bm\sigma$ can be visualized by considering a 3D vector $\vec{\sigma}$ = ($\sigma_{xx}$, $\sigma_{yy}$, $\sqrt{2}\sigma_{xy}$).
This is because the magnitude of stress, as given by the Frobenius norm,
\begin{equation}
\sigma = \| {\bm\sigma} \| = \sqrt{\sigma_{xx}^2 + \sigma_{yy}^2 + 2\sigma_{xy}^2} \, ,
\label{eqn:stress_magnitude}
\end{equation}
is the same as the length of the vector $\vec{\sigma}$.
\revision{Accordingly, in this work, we define the stress space of the plane stress condition by three axes: $\sigma_{xx}$, $\sigma_{yy}$, and $\sqrt{2}\sigma_{xy}$.  Every plane stress state corresponds to a vector $\vec{\sigma}$ (from the origin to a given point) in this 3D stress space.  The magnitude of the stress corresponds to the length of this vector.  Furthermore,} for every plane stress state with non-zero magnitude, we can define a unit vector, $\hat{\sigma} = \vec{\sigma} / \sigma$, which specifies the stress orientation.
Hence, all stress orientations in plane stress correspond to a unit sphere in the 3D space of ($\sigma_{xx}$, $\sigma_{yy}$, $\sqrt{2}\sigma_{xy}$).

\begin{figure}[h!]
    \centering
    \includegraphics[width=0.6\textwidth]{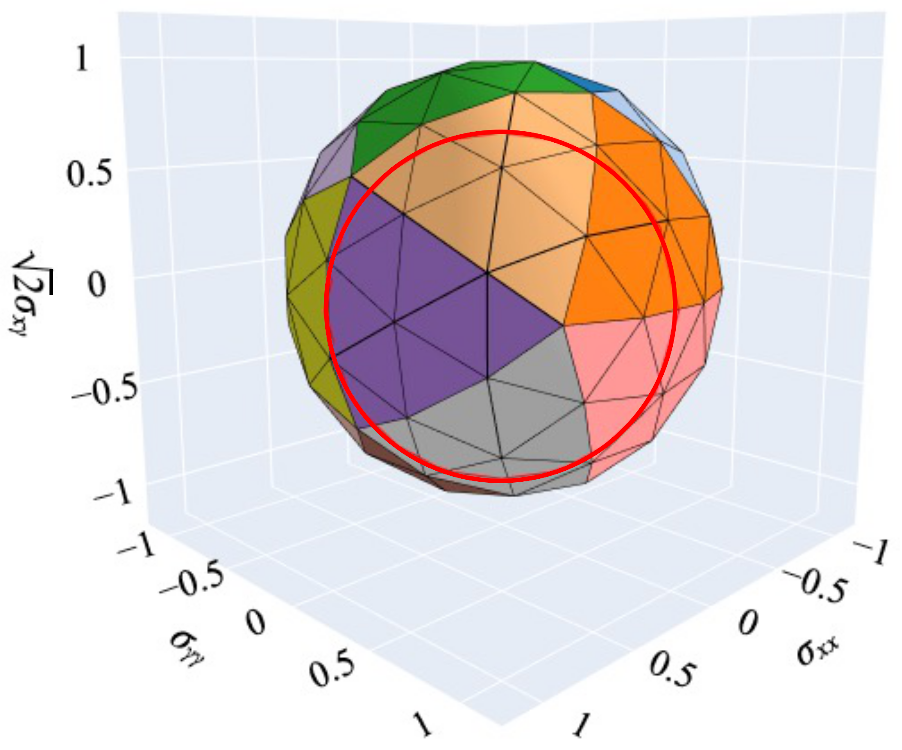}
    \caption{A total of 92 points distributed nearly uniformly on the unit sphere are shown as vertices of a polyhedron generated by the icosphere module in Python \citep{icosphere2023}. The red circle corresponds to uniaxial tensile loading along different directions in the $x$-$y$ plane.
    }
    \label{fig:plane_stress_state}
\end{figure}

\revision{We emphasize that the physical space specified by $(x, y, z)$ and the stress space specified by ($\sigma_{xx}$, $\sigma_{yy}$, $\sqrt{2}\sigma_{xy}$) are two different concepts, which nonetheless have some connections if we consider uniaxial loading conditions.}
\revision{For every uniaxial loading condition within the $x$-$y$ plane, whose loading axis can be specified by a vector in the physical space, there is a corresponding stress orientation vector in the stress space. For example, for uniaxial loading along the $[100]$ orientation (i.e. $x$-axis), the only non-zero stress component is $\sigma_{xx}$, so the corresponding vector in the stress space points along the $\sigma_{xx}$ axes. In fact, the stress orientations of all the uniaxial tensile loading conditions (with loading axis in the $x$-$y$ plane) correspond to a circle on the unit sphere in the stress space, as shown in \autoref{fig:plane_stress_state}.}
\revision{However, the uniaxial loading condition is only a subset of the plane stress condition; this means that most of the stress orientations in the ($\sigma_{xx}$, $\sigma_{yy}$, $\sqrt{2}\sigma_{xy}$) space do not correspond to uniaxial loading and hence do not have a corresponding ``loading direction'' in the physical space.}
%

To construct the yield surface under plane stress \revision{loading}, we thus need to sample the stress orientation $\hat{\sigma}$ from \revision{the} unit sphere \revision{in the 3D stress space of ($\sigma_{xx}$, $\sigma_{yy}$, $\sqrt{2}\sigma_{xy}$)}, and for each chosen $\hat{\sigma}$ increase the stress magnitude $\| {\bm\sigma} \|$ linearly with time until yielding is detected.
Our initial sampling of the plane stress space corresponds to 92 points nearly uniformly distributed on the unit sphere, which are the vertices of the polyhedron shown in \autoref{fig:plane_stress_state}. 
The coordinates of the sampling points are obtained using the icosphere module in Python \citep{icosphere2023}.
The uniaxial loading conditions along different directions in the $x$-$y$ plane are shown as the red circle on the unit sphere in the 3D space of plane stress.

For each stress orientation shown in \autoref{fig:plane_stress_state}, we performed a DDD simulation in which the stress orientation remains fixed and the stress magnitude increases at a constant rate $\dot\sigma = 10^{13} \,{\rm Pa}\cdot{\rm s}^{-1}$.
The magnitude of the stress rate is chosen here so that the stress-strain response of the single crystal \revision{Cu} \revision{under the} uniaxial loading condition \revision{at this stress rate} matches \revision{that} \revision{under} a strain rate of $10^3 \, {\rm s}^{-1}$, which is a common strain rate used in DDD simulations of single crystal Cu~\citep{sills2018prl,bertin2019msmse,akhondzadeh2020jmps,akhondzadeh2021mt}.
For example, \autoref{fig:stress_strain_curves_comparison} shows that for uniaxial tensile loading along $[100]$ crystal orientation, the stress-strain curves for constant stress rate of $10^{13} \, {\rm Pa}\cdot{\rm s}^{-1}$ and constant strain rate of $10^3 \, {\rm s}^{-1}$ are consistent with each other up to the yielding point.

\begin{figure}[h!]
    \centering
    \includegraphics[width=1.0\textwidth]{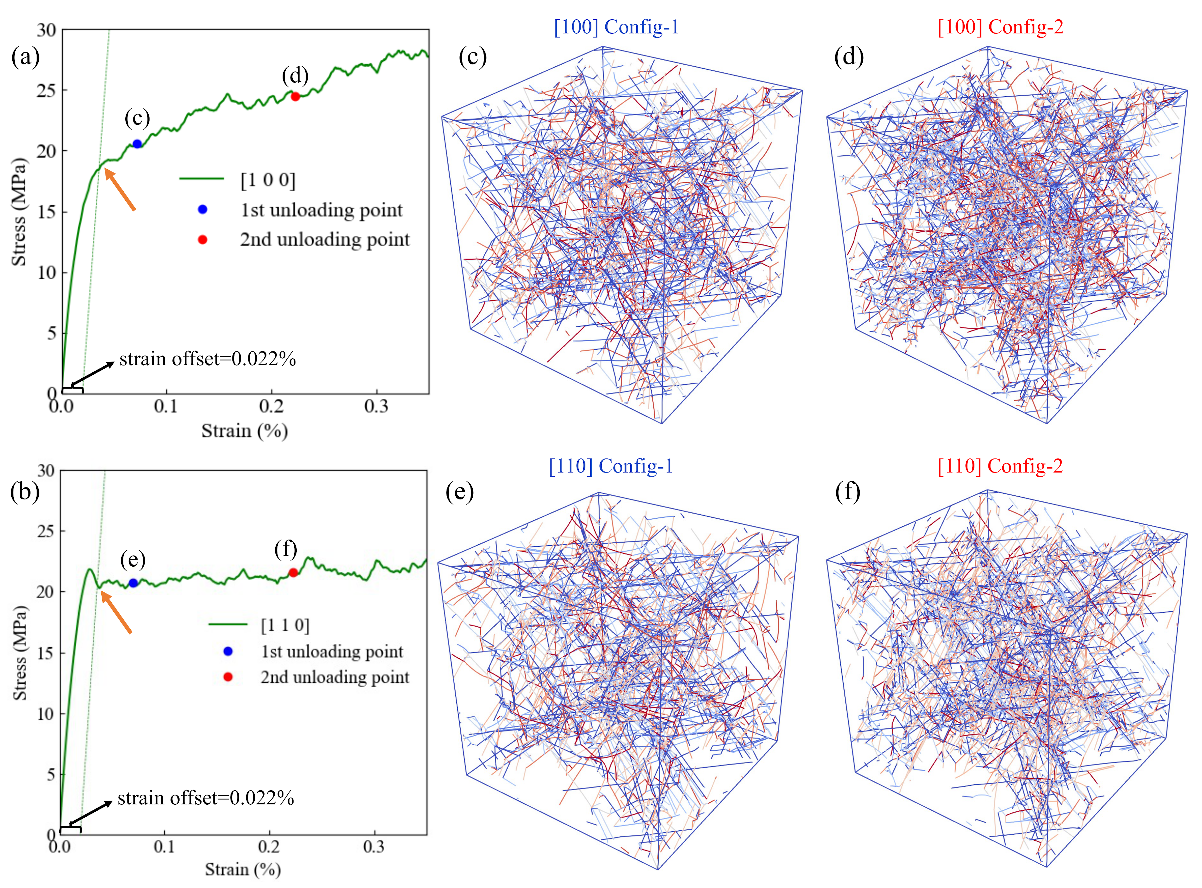}
    \caption{(a-b) The stress-strain curves for the uniaxial tensile loading at strain rate $10^{3}\,{\rm s}^{-1}$ along [100] and [110] directions, respectively. The blue and red dots on the curves represent the unloading points at the strains of 0.07\% and 0.22\%, where dislocation configurations are extracted and relaxed, respectively. The yield point is determined as the intersection of the curve with the elastic curve offset of 0.022\%, as denoted by the orange arrows. (c-d) show the corresponding relaxed dislocation configurations from the [100] loading, respectively. (e-f) show the corresponding relaxed dislocation configurations corresponding from the [110] loading curve, respectively.}
    \label{fig:uniaxial_loading_curves}
\end{figure}

\revision{To determine the yield surface from DDD simulations, we need to decide whether an appreciable amount of plastic strain has been accumulated as we increase the stress magnitude while keeping the stress orientation fixed.}
\revision{Given that both stress (divided by its magnitude) and plastic strain are tensors, the dot product between the two is a reasonable choice as a measure of the onset of yielding under the specified stress orientation, given by}
\begin{equation}
  \varepsilon^{\rm pl} = \hat{\sigma}_{xx}\varepsilon_{xx}^{\rm pl} + \hat{\sigma}_{yy}\varepsilon_{yy}^{\rm pl} 
              + 2 \hat{\sigma}_{xy}\varepsilon_{xy}^{\rm pl}
\end{equation}
where $\hat{\sigma}_{xx} = \sigma_{xx}/\|{\bm\sigma}\|$, $\hat{\sigma}_{yy} = \sigma_{yy}/\|{\bm\sigma}\|$, $\hat{\sigma}_{xy} = \sigma_{xy}/\|{\bm\sigma}\|$. 
$\varepsilon_{xx}^{\rm pl}$, $\varepsilon_{yy}^{\rm pl}$ and $\varepsilon_{xy}^{\rm pl}$ are the plastic strain components. 
This definition reduces to the normal plastic strain along the loading direction for the case of uniaxial loading.
In this work, the yield point of plane stress loading is determined by $\varepsilon^{\rm pl} = 0.022\%$.
This yield criterion is the same as using the strain offset in the case of uniaxial loading when applied to the stress-strain curve in which the strain is the total strain, as shown in \autoref{fig:uniaxial_loading_curves}(a) and (b). 
The choice of $0.022\%$ offset strain is similar to those used in previous experimental work ~\citep{franciosi1980acta}.
\revision{More justifications of this choice are provided below.}

To investigate how the yield surface evolves by strain hardening, we also computed the yield points along the stress orientations defined in \autoref{fig:plane_stress_state} for four more dislocation configurations.
These configurations are obtained from DDD simulations starting from \revision{Config-0} as the initial condition and subjected to uniaxial tensile loading conditions with strain rate $10^{3}\,{\rm s}^{-1}$, along the $[100]$ and $[110]$ directions, respectively.
\revision{As shown in \autoref{fig:uniaxial_loading_curves}(a) and (b), the yield point is determined as the intersection of the curve with the elastic curve offset of 0.022\%, as denoted by the orange arrows. The $0.022\%$ offset strain is indeed smaller than the $0.2\%$ offset strain often used in the empirical definition of yield stress from experimentally measured stress-strain curves in tensile tests.  The larger offset strain $0.2\%$ is more reasonable for tests where the precision in the strain measurement is not very high.  In comparison, the strain increments at each time step of our DDD simulations is very small and on the order of $10^{-5}$.  Hence the DDD simulation corresponds to experiments in which the strain resolution is very high. From \autoref{fig:uniaxial_loading_curves}(a), it can be seen that between $0.022\%$ and $0.2\%$ strain, the stress-strain curve exhibits a noticeable amount of strain hardening. Given this observation, the choice of $0.022\%$ offset strain is more reasonable than $0.2\%$ offset strain as the definition of yield stress in this work. From \autoref{fig:uniaxial_loading_curves}(a) and (b), it can be seen that the yield point defined by the $0.022\%$ offset strain roughly corresponds to the location where the stress-strain curve exhibits the greatest ``bend'', where the slope transitions from the initial elastic regime (i.e. Young's modulus) to the plastic flow regime (i.e. strain hardening). Therefore, $0.022\%$ is chosen as the critical offset strain for determining the yield point in this work.}

For each loading orientation, two configurations are extracted at the strains of 0.07\% and 0.22\%, and unloaded to zero stress and relaxed, leading to \revision{Config-1} and \revision{Config-2}, respectively.
These four dislocation configurations are plotted in \autoref{fig:uniaxial_loading_curves}(c), (d), (e), and (f), respectively.
From the normal stress-strain curves in \autoref{fig:uniaxial_loading_curves}(a) and (b), it can be seen that the $[100]$ loading orientation has a higher strain hardening rate than the $[110]$ loading orientation.
This is because for $[100]$ loading, all four slip planes and 8 out of 12 slip systems are active, and their intersections lead to a high strain hardening rate.
In contrast, for $[110]$ loading, only two slip planes and four slip systems are active.
However, the stress-strain curves in \autoref{fig:uniaxial_loading_curves}(a) and (b) only \revision{describe how the flow stress along a single direction (in the stress-space) changes with plastic deformation during uniaxial loading along a given direction (in physical space). In other words, each stress-strain curve only describes how a single point on the yield surface evolves with plastic deformation.}
\revision{In contrast, the results in this work reveal much more than that; we show how a 3D cross section of the yield surface (i.e. the yield stress along all stress orientations under the plane-stress condition) evolves with plastic deformation (along a certain uniaxial loading direction).} 

\subsection{Construction of yield surface with geometric prior method}
\label{sec:2.2}

\begin{figure}[h!]
    \centering
    \includegraphics[width=0.75\textwidth, trim=5 5 5 5,clip]{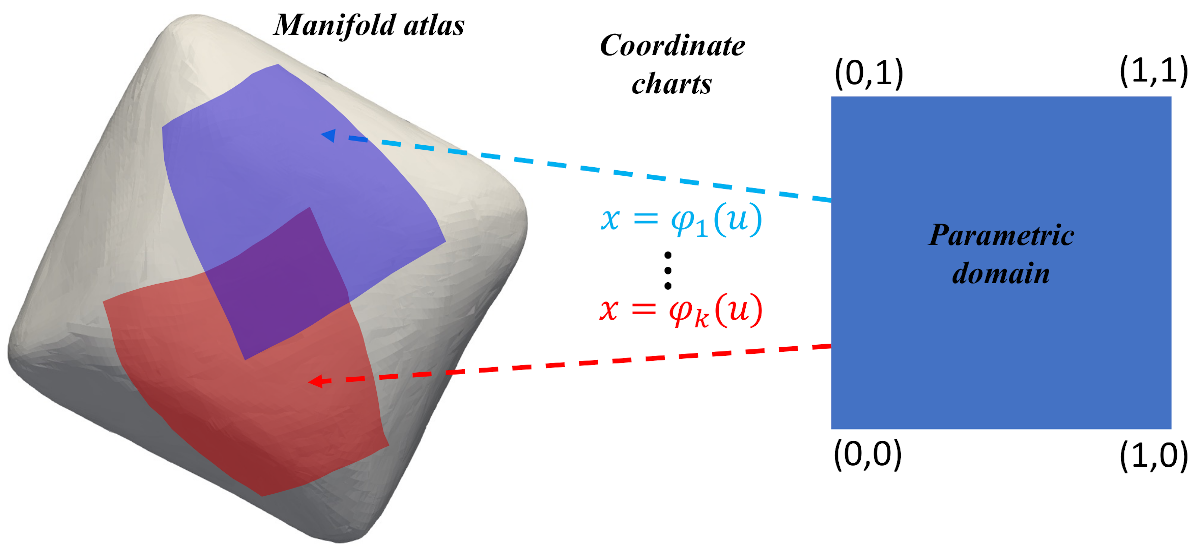}
    \caption{Sketch of the geometric prior manifold representation. The Riemann surface is divided into multiple pieces (patches), and each patch is associated with a coordinate chart as a mapping from a regular parametric domain.}
    \label{fig:geom-priors}
\end{figure}

We use the method of geometric prior~\citep{xiao2022cmame} to construct the yield surface ($\mathcal{S}$) under plane stress as a manifold atlas in $\mathbb{R}^3$, based on the yield point data predicted by DDD simulations along a discrete set of stress orientations ($\mathcal{X}$).
The method of geometric prior also allows us to iteratively update the yield surface by leveraging the geometrical interpretation of the yield surface. In particular, we calculate the curvature of the yield surface to the patch(es), at which denser data could be needed to capture the rapid change of plastic flow of the yield surface due to perturbation in the stress path. 
We characterize the mathematical representation of the yield surface as a smooth manifold atlas rather than the traditional implicit yield function representation \citep{hill1998mathematical}. One can draw an analogy between a manifold atlas and a piece-wise function: (1) the domain of the piece-wise function is divided into multiple pieces; similarly, the manifold atlas also partitions the surface into multiple patches. (2) On each piece of the domain, the piece-wise function is defined continuously, either explicitly or implicitly; correspondingly, each patch of the manifold atlas is associated with a coordinate chart, a neural network-trained nonlinear mapping that enables us to provide local coordinates of the patch within the yielding manifold. An example of this representation is shown in \autoref{fig:geom-priors}.

\begin{figure}[h!]
    \centering
    \includegraphics[width=0.75\textwidth]{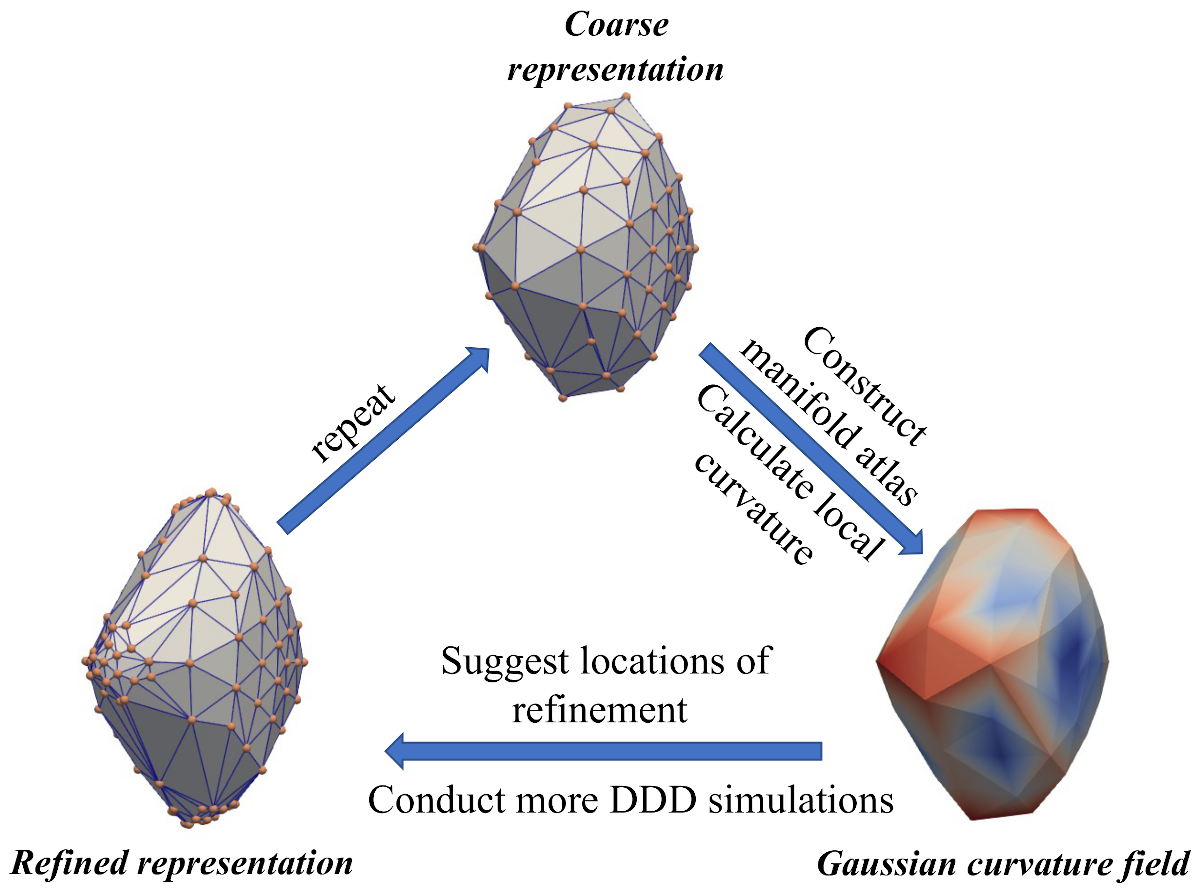}
    \caption{Major steps in the reconstruction of a yield surface.}
    \label{fig:ML-big-picture}
\end{figure}

After constructing the initial manifold atlas based on DDD simulation data of 92 stress orientations, we can compute local geometric information, e.g., Gaussian curvature, everywhere on the yield surface. We then identify regions where local refinement is desirable.  These are regions where the yield surface intersects a bounding sphere of a chosen radius and is centered at points of local maxima of Gaussian curvature.  A point cloud is generated in each of these refinement regions using Poisson disk sampling.  Additional DDD simulations are then performed to predict yield points at stress orientations corresponding to these new sampling points.  The data is combined with the previously obtained DDD data to obtain a refined manifold atlas representing the updated yield surface.  The process can be iterated to sequentially refine the yield surface representation, as illustrated in \autoref{fig:ML-big-picture}.

\subsubsection{Geometric prior method for yield surface construction}
\label{sec:2.2.1}
Here, we provide some specifics of the geometric prior method that is applied to constructing the yield surface $\mathcal{S}$ under plane stress as a manifold atlas in $\mathbb{R}^3$; for more details, please refer to~\citep{xiao2022cmame}.
To generate the patch discretization of $\mathcal{S}$, we prescribe a set of anchor points $\Vec{x}_p$ on $\mathcal{S}$. Then, a surface patch $\mathcal{S}_p$ is defined by the intersection of the neighborhood ball at $\Vec{x}_p$ and $\mathcal{S}$: $\mathcal{S}_p = \mathcal{S} \cap \mathcal{B}_p, \mathcal{B}_p = \{ \Vec{x}\in \mathbb{R}^3 |\:\: ||\Vec{x} - \Vec{x}_p\| < \epsilon \}$, where $\epsilon$ is the radius of the neighborhood ball and $\|\cdot\|$ is the L2-norm. We then define the parametrization of the local coordinate charts as $\phi_p(\vec{v}): V \rightarrow \mathbb{R}^3$, where $V=(0,1)\times (0,1)$. 

Subsequently, there are two major learning tasks: (1) fit the local coordinate charts $\phi_p(\Vec{v})$; (2) ensure the consistency between patches. The first task is achieved by approximating each coordinate chart with a multi-layer-perceptron (MLP) function \citep{murtagh1991multilayer} and optimizing the learnable parameters with the Sinkhorn regularized distance \citep{cuturi2013anip}:
\begin{equation}
\mathcal{L}_p = \min_{P_{ij}}\: \sum_{i,j\leq N_p} P_{ij} \| \phi(\vec{v}_j; \tensor{W}_p) - \vec{x}_j \|^2 + \chi^{-1} \sum_{i,j\leq N_p} P_{ij}\log P_{ij}
\label{eqn:sinkhorn_local_chart}
\end{equation}
where $\mathcal{L}_p$ is the training loss function and $\tensor{W}_p$ is the set of neural network parameters for the $p$th coordinate chart, such that $\phi_p(\Vec{v}) \approx \phi(\Vec{v}; \tensor{W}_p)$.
$N_{p}$ is the number of sampling points used to calculate the Sinkhorn distance, which is less than or equal to the total available points within the same coordinate chart. 
$\vec{v}_j\in V$ indicates the input samples following the Poisson disk distribution for the reconstructed set. $P_{ij}$ is an $n \times n$ bi-stochastic matrix. $\chi$ is a regularization parameter such that $\mathcal{L}_p$ approximates the optimal transport distance \citep{rubner1997earth} as $\chi \rightarrow \infty$.

The second task is performed in two steps: we first find the optimal indices permutation policy of each patch $p$ via the following objective derived from \eqref{eqn:sinkhorn_local_chart}:
\begin{equation}
\min_{\tensor{W}_p}\: \inf_{\pi_p}
\sum_{i\leq N_p} \|\phi(\vec{v}_{i}; \tensor{W}_p) - x_{\pi_p(i)} \|^2
\label{eqn:global_loss_1}
\end{equation}
where $\pi_p$ is a permutation policy, assigning indices of points in $\mathcal{X}_p$ to indices of parametric positions in $V_p$.
We then minimize the divergence between embedding functions $\phi_p, \phi_q$ for all pairs of overlapping patches:
\begin{equation}
\min_{\tensor{W}_p, \tensor{W}_q}\: \inf_{\pi_p, \pi_q}\:\:
\sum_{i\in T_{pq}} \|\phi(\vec{v}_{i}; \tensor{W}_p) - \phi(\vec{v}_{\pi^{-1}_q(\pi_p(i))}; \tensor{W}_q) \|^2
\label{eqn:global_loss_2}
\end{equation}
where $T_{pq} = \{ i \, | \, \vec{x}_{\pi_p(i)} \in \mathcal{X}_p \cap \mathcal{X}_q \}$ indicates the set of indices of parametric points in chart $p$ included within the intersection of chart $p$ and $q$.

\subsubsection{Local refinement of yield surface representation}
\label{sec:2.2.2}

The initial (nearly uniform) sampling of the plane stress conditions provides an overall representation of the yield surface that may be too coarse.
Because of the substantial computational cost of DDD simulations, it is preferable to locally refine the yield surface representation in regions with sharp edges or corners than a uniform refinement.
Similar to mesh refinement \citep{berger1984jcp} in numerical simulations, the point cloud (for which DDD predictions are collected) should be densified at the locations where it is likely to provide the most information gain. 
In geometry reconstruction, we believe curvature is a reasonable choice of quantitative criterion for such purpose \citep{tang2018mg}.

There are multiple curvature measures in the application of geometric learning. In order to efficiently establish a criterion with inequality, we adopted one of the scalar curvature measures: Gaussian curvature \citep{kreyszig2013differential}, denoted as $\mathcal{K}$ (for mathematical formulation of Gaussian curvature, please refer to \citep{xiao2022cmame}). Gaussian curvature can serve as a smoothness measure for digital images and geometric objects \citep{tang2023jsps}, where a large Gaussian curvature usually indicates a relatively abrupt change in the geometric shape (e.g. non-smoothness).
Here we selected regions for enrichment as locations where $\mathcal{K} > 0.07 \mathrm{~MPa}^{-2}$.
As \autoref{fig:ML-big-picture} shows, such locations group together around the ``\textit{corners}'' of the yield surface.
For each of the corners, we then defined a bounding sphere $\mathbf{\mathcal{B}}^{(e)}$ with minimum radius $R^{(e)}$ that includes all the selected data points (whose $\mathcal{K} > 0.07 \mathrm{~MPa}^{-2}$), where the superscript $e$ labels the enrichment regions (corners).

To refine the yield surface inside the bounding sphere $\mathbf{\mathcal{B}}^{(e)}$, we first projected the bounding sphere onto the unit sphere $\tilde{S}^2$ centered at the origin.  The projection is defined as $\mathcal{P}_{\tilde{S}^2}(\mathbf{\mathcal{B}}^{(e)})$.
We used Poisson disk sampling~\citep{bowers2010atg, point-cloud-utils} to generate a random but relatively uniform distribution of points over the entire unit sphere $\tilde{S}^2$ and extracted sampling points that lie within $\mathcal{P}_{\tilde{S}^2}(\mathbf{\mathcal{B}}^{(e)})$.
Each one of these sampling points on the unit sphere represents a stress orientation $\hat{\sigma}$ for which a DDD simulation is performed under constant stress rate loading to determine the yield point.

\begin{figure}[h!]
    \centering
    \includegraphics[width=0.5\textwidth]{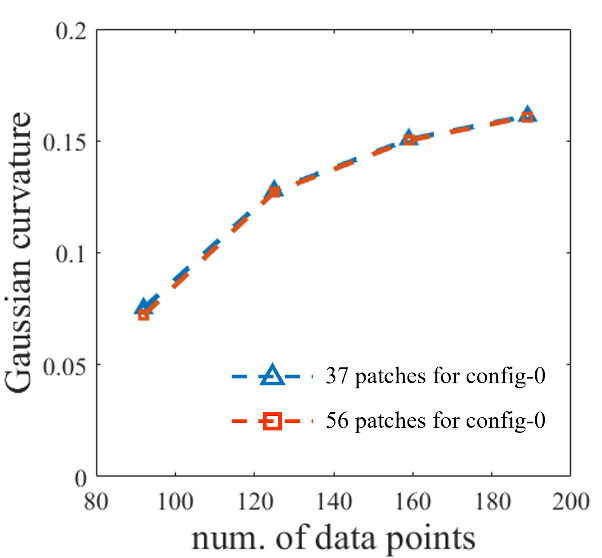}
    \caption{Gaussian curvature of the learned yield surface manifold for \revision{C}onfig-0 at the stress orientation $\hat{\sigma} = \left(-\frac{\sqrt{2}}{2}, -\frac{\sqrt{2}}{2}, 0\right)$ as a function of the total number of stress orientations used to construct the yield surface manifold.
    \revision{The results from using 37 patches and 56 patches in the geometric prior method are plotted.}}
    \label{fig:gauss-curve-refine}
\end{figure}

To determine how many sampling points are needed in the refined regions, we monitor the convergence of the local Gaussian curvature predicted by the geometric prior at a given stress orientation, $\hat{\sigma} = \left(-\frac{\sqrt{2}}{2}, -\frac{\sqrt{2}}{2}, 0\right)$.
\autoref{fig:gauss-curve-refine} plots the predicted Gaussian curvature at this point when 
33, 67 and 97 additional data points are combined with the initial 92 data points to construct the yield surface manifold.
We observe a gradual increase of the predicted local Gaussian curvature and an indication of convergence.  Due to the limitations of computational resources, we limit the total number of sampling points on the unit sphere to 189, with each sampling point corresponding to a DDD simulation at a different stress orientation.
\revision{To demonstrate that the estimated Gaussian curvature is insensitive to the number of patches used to represent the yield surface in the geometric prior method, Figure~\ref{fig:gauss-curve-refine} plots the estimates from using 37 patches and 56 patches, respectively.  The effect of the number of patches appears negligible.  For the remaining part of this paper, we will use 37 patches to represent the yield surface.}

\begin{figure}[h!]
    \centering
    \includegraphics[width=0.7\textwidth]{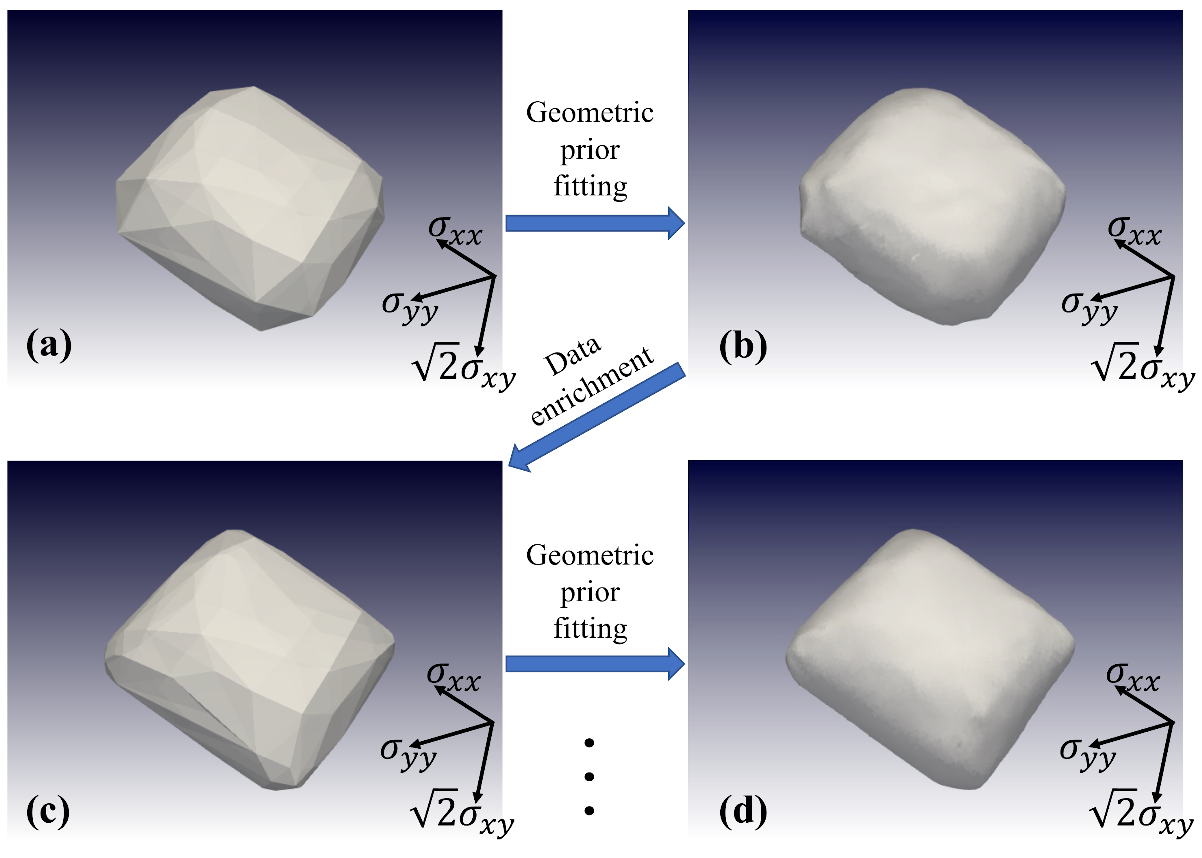}
    \caption{The geometric refinement of yield surface for \revision{Config}-0. (a) The yield surface is generated by interpolated triangulation from the initial 92 data points of yielding stress. (b) Based on (a), the yield surface is constructed by the fitting of geometric prior. (c) Based on (b), more sampling points are selected where DDD simulations are performed for data enrichment, resulting in 189 data points on the yield surface. (d) Based on (c) the yield surface is constructed by the fitting of geometric prior.}
    \label{fig:refining-configs}
\end{figure}

\revision{An active learning process is then applied to enrich the dataset and improve the accuracy of the DDD-predicted yield surface at selected locations.} 
This iterative procedure is illustrated in \autoref{fig:refining-configs}.
The original and refined constructions of the yield surface for \revision{Config-0} are shown in \autoref{fig:refining-configs}(b) and (d), respectively.
Limited by the computing power and the high computational cost of DDD simulations, the data enrichment step cannot be continued beyond the first iteration, and we cannot obtain thousands of data points, a typical number of training data points for surface reconstruction \citep{williams2019deep}.
In this work, we perform one step of data enrichment, and our final yield surface is trained on 189 data points from DDD simulations.

\section{Results}
\label{sec:results}
\subsection{Yield surfaces from plane stress loading}
\label{sec:3.1}

\begin{figure}[h!]
    \centering
    \includegraphics[width=0.95\textwidth]{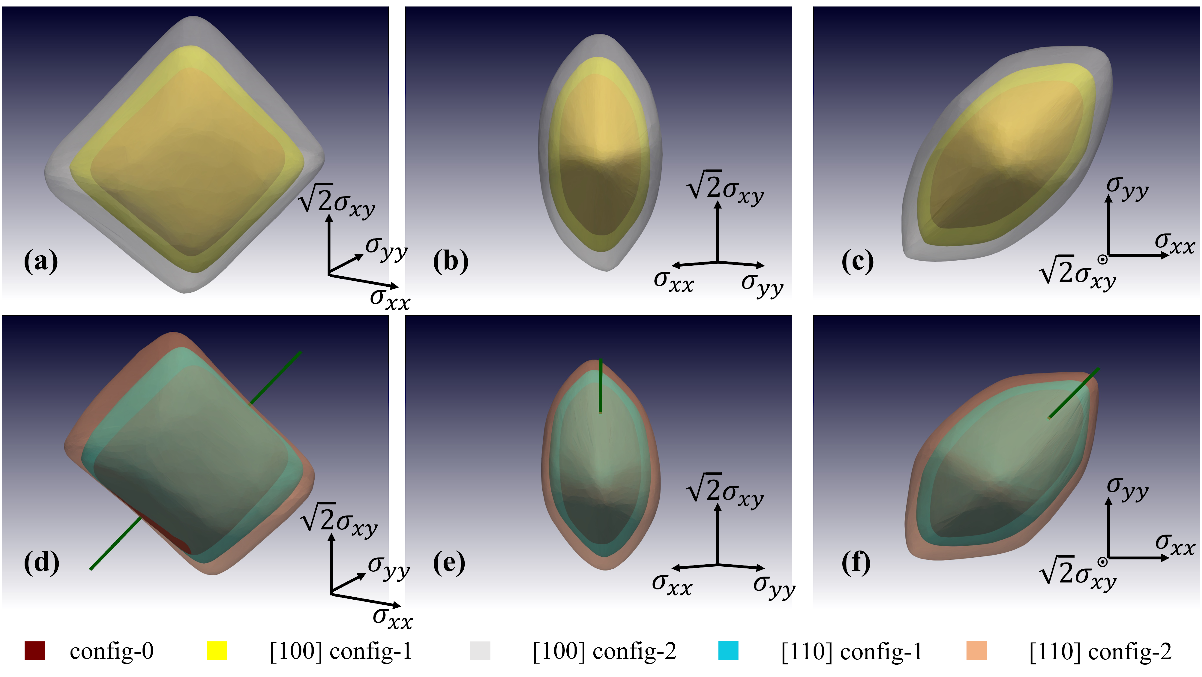}
    \caption{Along three different viewing angles, yield surfaces of (a-c) \revision{Config}-0, [100] \revision{Config}-1 and [100] \revision{Config}-2, and (d-f) \revision{Config}-0, [110] \revision{Config}-1 and [110] \revision{Config}-2 under plane stress loading. The green straight line in (d-f) denotes the stress orientation of uniaxial loading along [110] crystal orientation, i.e., $\hat{\sigma} = \left( 1/2, 1/2, \sqrt{2}/2\right)$.}
    \label{fig:five_YS} 
\end{figure}

\autoref{fig:five_YS} shows the yield surface constructed from the DDD simulations under plane stress loading from the initial dislocation configuration (\revision{Config}-0), together with how the yield surface evolves by plastic deformation along [100] and [110] directions, respectively.
\autoref{fig:five_YS} (a-c) show the yield surfaces from different viewing angles for \revision{Config}-0, [100] \revision{Config}-1, [100] \revision{Config}-2. It can be seen that with increasing plastic strain along [100] crystal orientation, the yield surface appears to expand isotropically in all stress orientations. This indicates that plastic deformation along [100] direction results in \textit{isotropic hardening}.
\autoref{fig:five_YS} (d-f) show the yield surfaces from different viewing angles for \revision{Config}-0, [110] \revision{Config}-1, [110] \revision{Config}-2. It can be seen that with increasing plastic strain along [110] crystal orientation, the expansion of the yield surface is not isotropic in all stress directions. Specifically, the yield surface appears to be not expanding appreciably along the stress orientations neighboring that of the [110] uniaxial loading itself.  (The stress orientation corresponding to [110] uniaxial loading is indicated by the green straight lines.) This is consistent with the low strain-hardening rate in the stress-strain curve of [110] loading shown in \autoref{fig:uniaxial_loading_curves}(b).
However, in stress orientations ``perpendicular'' to that of the [110] uniaxial loading in the space of $\left(\sigma_{xx}, \, \sigma_{yy}, \sqrt{2}\sigma_{xy}\right)$, the expansion of yield surface (i.e. strain hardening) is significant. This indicates that plastic deformation along [110] direction results in \textit{latent hardening}, representing the phenomenon that yield stress and hardening rate are higher in previously unactivated slip systems than in the previously activated primary slip systems \citep{edwards1953tmsa,nakada1969pssb,kocks2003pms}. Latent hardening has been reported in the experiments of FCC single crystals where the initial loading orientations are single-slip orientations, which lie in the central region of the stereographic triangle \citep{kocks1966acta,jackson1967cjp,wessels1969acta}.
In comparison, the $[110]$ loading direction is at a corner of the stereographic triangle and activates multiple slip systems.

\begin{figure}[h!]
    \centering
    \includegraphics[width=\textwidth]{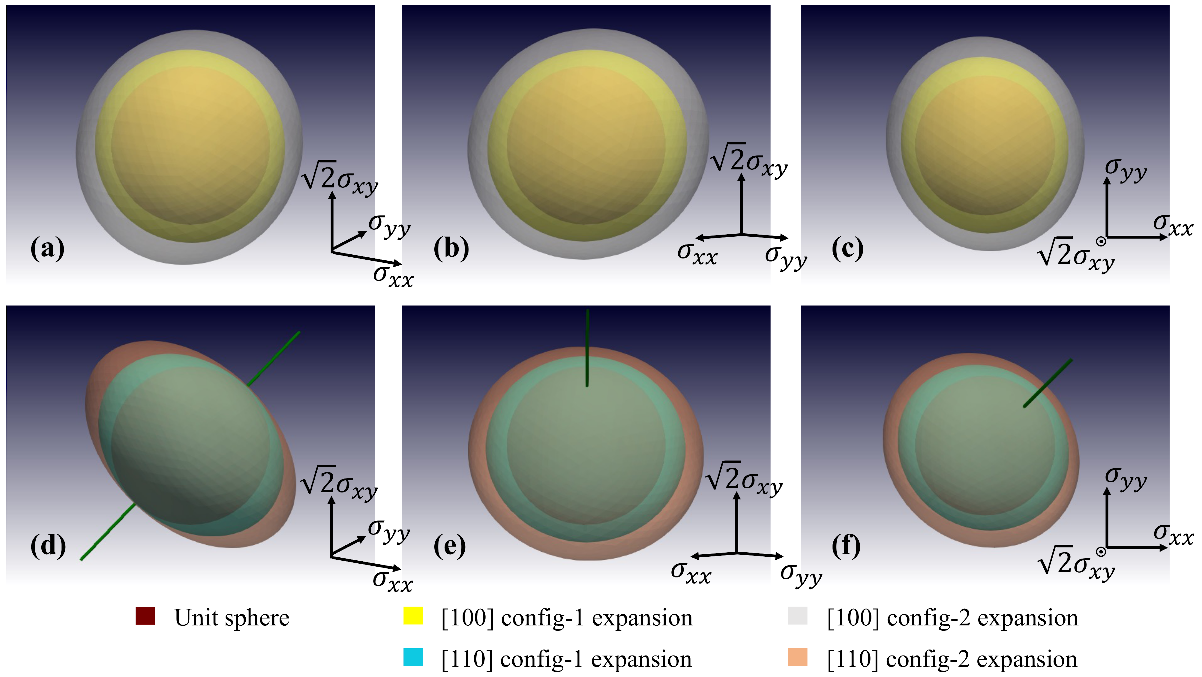}
    \caption{Hardening ratios of (a)-(c) [100] \revision{Config}-1 and [100] \revision{Config}-2, and (d)-(f) [110] \revision{Config}-1 and [110] \revision{Config}-2 in different perspectives. The innermost unit sphere is plotted as a reference. The stress orientation $\hat{\sigma} = \left({1}/{2}, {1}/{2}, {\sqrt{2}}/{2}\right)$ of the [110] uniaxial loading is shown as a green line.}
    \label{fig:hardening_rate}
\end{figure}

To better illustrate the strain hardening behaviors, in \autoref{fig:hardening_rate} we plot the hardening ratio, which is defined as the yield surfaces at different plastic strains over the initial yield surface (i.e. \revision{Config}-0).
(The innermost shape is a unit sphere plotted as a reference.)
For each configuration, the orientation-dependent hardening ratio data is fitted to an ellipsoid for smoothing.
\autoref{fig:hardening_rate} (a-c) shows that for plastic deformation along [100] crystal orientation, the hardening ratio is nearly isotropic in all stress orientations.
In contrast, \autoref{fig:hardening_rate} (d-f) shows that for plastic deformation along [110] crystal orientation, the hardening ratio is close to 1 (i.e. no hardening) along the stress orientation corresponding to the [110] loading orientation itself, and is greater than one and nearly isotropic in the stress orientations perpendicular (in the $\hat{\sigma}$ space) to that of [110] uniaxial loading.

\begin{figure}[h!]
    \centering
    \includegraphics[width=0.95\textwidth]{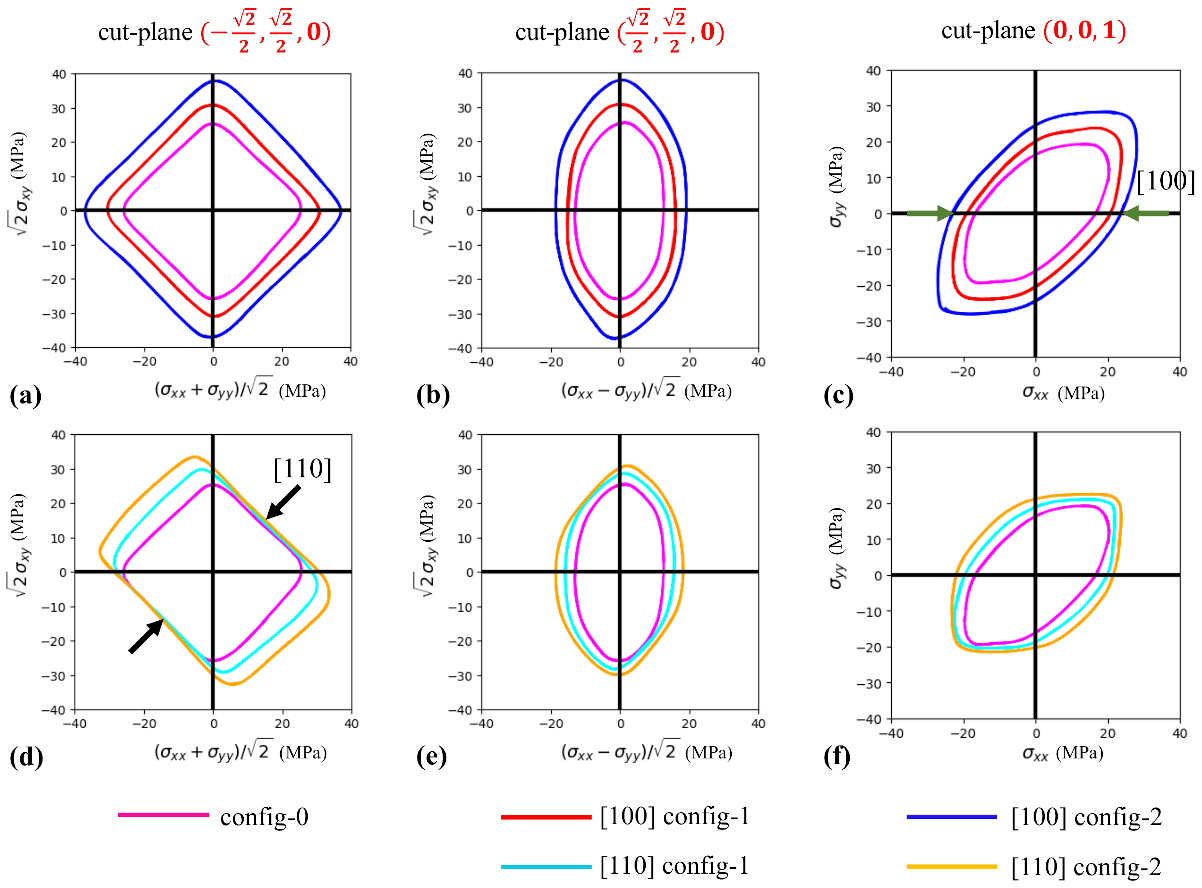}
    \caption{Cut-plane views of the yield surfaces of (a-c) \revision{Config}-0, [100] \revision{Config}-1 and [100] \revision{Config}-2, and (d-f) \revision{Config}-0, [110] \revision{Config}-1 and [110] \revision{Config}-2 under plane stress loading. The green and black arrows denote the stress orientations of [100] and [110] uniaxial loadings, respectively.}
    \label{fig:yld-surf-cutplane-all}
\end{figure}

For a more detailed examination of how the yield surface evolves with plastic strain, \autoref{fig:yld-surf-cutplane-all} plots the cross-section views of the yield surfaces for different cut-planes in the space of stress orientations.
When viewed on the cut-plane $\left(-\frac{\sqrt{2}}{2}, \frac{\sqrt{2}}{2}, 0\right)$, the yield surface for the original configuration (\revision{Config}-0) has approximately a square shape.
\autoref{fig:yld-surf-cutplane-all}(a) shows that with [100] uniaxial loading, the cross-section view of the yield surface expands while keeping the square shape.
In contrast, \autoref{fig:yld-surf-cutplane-all}(d) shows that with [110] uniaxial loading, the cross-section view of the yield surface expands asymmetrically into a rectangular shape.
\autoref{fig:yld-surf-cutplane-all}(b-c) shows the cross-section views of yield surfaces under [100] uniaxial loading on cut-planes perpendicular to that in (a); the yield surface expands nearly isotropically in all directions.
In comparison, \autoref{fig:yld-surf-cutplane-all}(e-f) shows corresponding views of yield surfaces under [110] uniaxial loading, where the expansion appears less in magnitude and is asymmetric.

\subsection{Yield loci for uniaxial tension}
\label{sec:3.2}

Experimentally uniaxial loading tests are much easier to carry out than the general plane-stress loading (which may be performed on a tube with combined tension and torsion \citep{taylor1931ptrsa}).
Therefore, we examine the predicted yield conditions under uniaxial tension. The uniaxial tensile loading conditions correspond to a sub-set of stress orientations within the plane-stress conditions and are shown as the red circle on the unit sphere in \autoref{fig:plane_stress_state}. 
Note that the center of the circle is not at the origin in the $\hat{\sigma}$ space. The central inversion of this circle with respect to the origin produces another circle, which corresponds to uniaxial compressive loading conditions.

\begin{figure}[h!]
    \centering
    \includegraphics[width=0.9\textwidth]{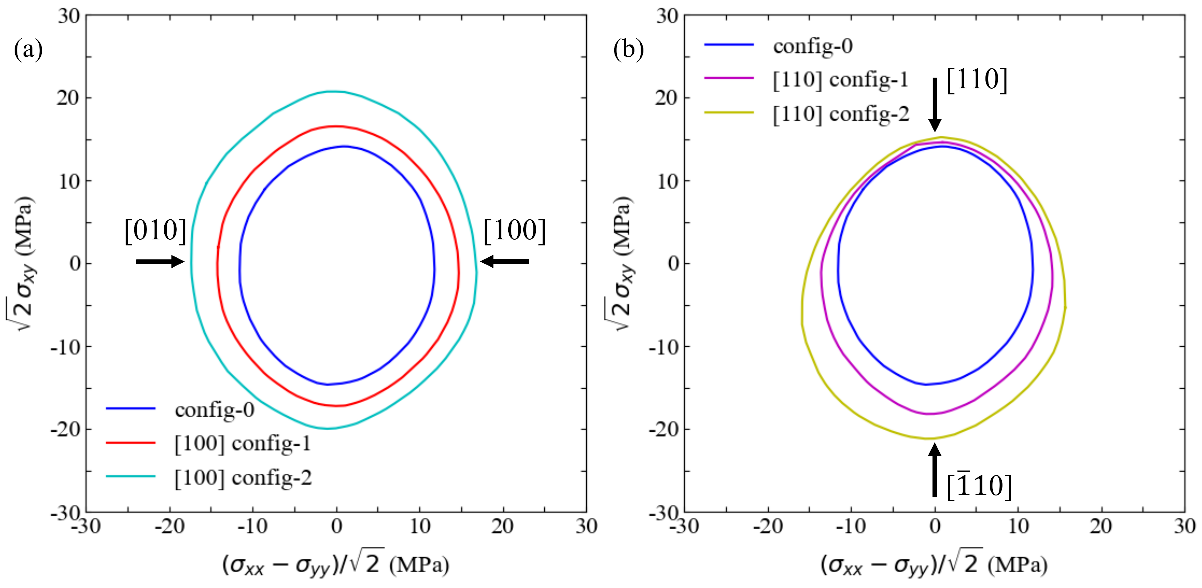}
    \caption{The 2D view of the yield loci along the $\hat{\sigma} = \left(\sqrt{2}/2,\sqrt{2}/2,0\right)$ direction for uniaxial tensile loading applied to (a) \revision{Config}-0, [100] \revision{Config}-1 and [100] \revision{Config}-2, and (b) \revision{Config}-0, [110] \revision{Config}-1 and [110] \revision{Config}-2. 
    }
    \label{fig:uniaxial_hardening}
\end{figure}

\autoref{fig:uniaxial_hardening} plots the yield loci for the uniaxial tensile loading conditions, in terms of $\sqrt{2}\sigma_{xy}$ v.s. $(\sigma_{xx}-\sigma_{yy})/\sqrt{2}$.
This figure corresponds to viewing the yield loci corresponding to uniaxial tension on the yield surface along the $\hat{\sigma} = \left(\sqrt{2}/2,\sqrt{2}/2,0\right)$ direction.
\autoref{fig:uniaxial_hardening}(a) shows the evolution of yield loci from \revision{Config}-0 during plastic deformation along the [100] direction.
The yield loci appear to expand isotropically while maintaining an oval shape.
We note that the Von Mises (i.e. $J_2$) yield criterion (widely used for polycrystals) would predict the yield loci (for uniaxial tension) in the shape of a circle when plotted using these axes \citep{mises1928jamm}.  Hence the deviation of the yield loci from a circle is a manifestation of the plastic anisotropy of single crystals.
\autoref{fig:uniaxial_hardening}(b) shows the evolution of yield loci from \revision{Config}-0 during plastic deformation along the [110] direction.
There is little expansion of the yield loci around the [110] loading direction itself. Nevertheless, a substantial expansion is observed in the opposite end of the loop, corresponding to the $[\bar{1}10]$ loading orientation.
The predictions of yield loci evolution from plastic deformation in the $[110]$ loading direction, especially the pronounced latent hardening in the $[\bar{1}10]$ direction, provide new insight into the strain hardening behavior of FCC crystals. 
Existing experimental studies on latent hardening in FCC crystals were usually conducted where the initial uniaxial loading is along a direction that favors single slip, i.e., only a single slip system is activated~\citep{franciosi1980acta,franciosi1985acta}. 
Subsequently, a ``daughter'' sample is cut from the ``parent'' sample and then loaded in a different orientation.
The goal was to let the primary slip system in the ``daughter'' sample interact strongly with the primary slip system previously activated in the ``parent'' sample to cause latent hardening.
We hope our prediction of the latent hardening behavior for initial loading along the $[110]$ direction will motivate new experiments for its verification.

\section{Discussion}
\label{sec:discuss}

\subsection{Strain hardening mechanisms}
\label{sec:4.1}

The flow stress $\tau$ of a metal is known to correlate with the dislocation density $\rho$.  In particular, the Taylor relation \citep{taylor1934prsla}, $\tau = \alpha \mu b \sqrt{\rho}$, is well supported by experimental data on polycrystals \citep{wang2021scripta}, where $\mu$ is the shear modulus, $b$ is the Burgers vector magnitude, and $\alpha$ is a dimensionless constant.
Here the total dislocation density for \revision{Config}-0 is $\sim 1.2 \times 10^{12} \mathrm{~m}^{-2}$.
After uniaxial tension along \revision{the} $[100]$ crystal orientation, the toal dislocation density becomes $\sim 1.3 \times 10^{12} \mathrm{~m}^{-2}$ and $\sim 2.0 \times 10^{12} \mathrm{~m}^{-2}$, for \revision{Config}-1 and \revision{Config}-2, respectively.
After uniaxial tension along \revision{the} $[110]$ crystal orientation, the total  dislocation density becomes $\sim 1.3 \times 10^{12} \mathrm{~m}^{-2}$ and $\sim 1.6 \times 10^{12} \mathrm{~m}^{-2}$, for \revision{Config}-1 and \revision{Config}-2, respectively.
Although in both deformation orientations, the strain hardening is accompanied by an increase of total dislocation density, this does not explain the difference between the isotropic hardening and latent hardening behaviors.
\revision{The Taylor relation has been generalized to relate the critical resolved shear stress on one (e.g. the dominant) slip system with the dislocation densities on every individual slip system~\citep{akhondzadeh2020jmps,franciosi1982acta}.}
\revision{In the following, we show that the isotropic hardening and latent hardening behaviors observed here can be qualitatively understood through the dislocation densities on individual slip systems and the interaction between slip systems.}

\begin{table}[htb] 
\centering
\small
\caption{Schmid factors and dislocation densities (in $10^{11} \mathrm{~m}^{-2}$) for the 12 slip systems in various dislocation configurations (Config-0, Config-1 and Config-2) extracted from the $[100]$ uniaxial loading. The active slip systems are highlighted in gray. Slip system index: $i$, Slip plane: \textbf{n}, Burgers vector: \textbf{b}, Schmid-Boas notation: \textbf{SB}, Schmid factors: \textbf{SF}} \label{tbl:dis_density_100}
\begin{tabular}{ P{1.0cm} P{1.2cm} P{1.2cm} P{1cm} P{1.2cm} P{1.5cm} P{1.5cm} P{1.5cm} }
\toprule
$i$ & \textbf{n} & \textbf{b} & \textbf{SB} & \textbf{SF} & Config-0 & Config-1 & Config-2 \\
\toprule
1  & $(\overline{1}11)$ & $\frac{1}{2}[0\overline{1}1]$ & A2 & 0 & 0.6945 & 0.6354 & {1.0221} \\
\rowcolor{lightgray} 2  & $(\overline{1}11)$ & $\frac{1}{2}[101]$ & A3 & 0.4082 & 0.8599 & {0.8833} & {1.4480} \\
\rowcolor{lightgray} 3  & $(\overline{1}11)$ & $\frac{1}{2}[110]$ & A6 & 0.4082 & 0.8882 & {1.0106} & {1.7339} \\
4  & $(111)$ & $\frac{1}{2}[0\overline{1}1]$ & B2 & 0 & 0.7711 & 0.5372 & {1.0167} \\
\rowcolor{lightgray} 5  & $(111)$ & $\frac{1}{2}[\overline{1}01]$ & B4 & 0.4082 & 0.6238 & {0.8580} & {1.3295} \\
\rowcolor{lightgray} 6  & $(111)$ & $\frac{1}{2}[\overline{1}10]$ & B5 & 0.4082 & 0.8691 & {1.2066} & {1.9012} \\
7  & $(\overline{1}\overline{1}1)$ & $\frac{1}{2}[011]$ & C1 & 0 & 0.8838 & 0.8195 & {1.0448} \\
\rowcolor{lightgray} 8  & $(\overline{1}\overline{1}1)$ & $\frac{1}{2}[101]$ & C3 & 0.4082 & 0.7518 & {0.8251} & {1.2895} \\
\rowcolor{lightgray} 9  & $(\overline{1}\overline{1}1)$ & $\frac{1}{2}[\overline{1}10]$ & C5 & 0.4082 & 0.8541 & 0.7792 & {1.1405} \\
10 & $(1\overline{1}1)$ & $\frac{1}{2}[011]$ & D1 & 0 & 0.7214 & 0.5979 & {0.8278} \\
\rowcolor{lightgray} 11 & $(1\overline{1}1)$ & $\frac{1}{2}[\overline{1}01]$ & D4 & 0.4082 & 0.7460 & {0.7744} & {1.4479} \\
\rowcolor{lightgray} 12 & $(1\overline{1}1)$ & $\frac{1}{2}[110]$ & D6 & 0.4082 & 0.6779 & 0.6584 & {0.9163} \\
\bottomrule 
\end{tabular}
\end{table}

\autoref{tbl:dis_density_100} shows the dislocation densities on individual slip systems during uniaxial tension along $[100]$ direction.
In FCC metals, there are four slip planes, each having three slip directions, with a total of 12 slip systems.
For uniaxial loading along $[100]$ direction, eight slip systems (2 on every slip plane) are activated with the same Schmid factor.  (The remaining 4 slip systems have zero Schmid factor.)
\autoref{tbl:dis_density_100} shows that the dislocation densities on the active slip systems increase with strain for both \revision{Config}-1 (strained to 0.07\%, except the slight decrease for C5 and D6 slip systems) and \revision{Config}-2 (strained to 0.22\%).
On the inactive slip systems (A2, B2, C1, D1), the dislocation densities in \revision{Config}-1 is actually lower than those in \revision{Config}-0; however, the dislocation densities on these slip systems in \revision{Config}-2 exceed those in \revision{Config}-0.
This phenomenon is termed ``slip-free multiplication''~\citep{akhondzadeh2021mt}, because dislocations multiply on these slip systems with zero Schmid-factor and zero shear strain rate.
Slip-free multiplication has been linked to co-planar interactions between active slip systems on the same slip plane (e.g. dislocations on inactive slip system A2 are produced because both A3 and A6 are active).
We conclude that the isotropic hardening produced by plastic deformation along $[100]$ is mainly because all 4 slip planes have 2 slip systems activated resulting in increased dislocation density on all slip planes. 
The slip-free multiplication is deemed not essential for the isotropy of strain hardening because even though dislocation densities on inactive slip systems (A2, B2, C1, D1) are lower in \revision{Config}-1 than in \revision{Config}-0, the yield surface of \revision{Config}-1 appears to have expanded isotropically from that of \revision{Config}-0 nonetheless.
In other words, having dislocations multiply on 2 out of 3 of slip systems on all 4 slip planes appears to be sufficient for isotropic hardening. Existing experiments also suggest that the difference of hardening behaviors in FCC single crystal metals depends on the activated slip planes instead of the slip directions or slip systems \citep{kocks1966acta}.

\begin{table}[htb] 
\centering
\small
\caption{Schmid factors and dislocation densities (in $10^{11} \mathrm{~m}^{-2}$) for the 12 slip systems in various dislocation configurations (Config-0, Config-1 and Config-2) extracted from the $[110]$ uniaxial loading. The active slip systems are highlighted in gray. Slip system index: $i$, Slip plane: \textbf{n}, Burgers vector: \textbf{b}, Schmid-Boas notation: \textbf{SB}, Schmid factors: \textbf{SF}} \label{tbl:dis_density_110}
\begin{tabular}{ P{1.0cm} P{1.2cm} P{1.2cm} P{1cm} P{1.2cm} P{1.5cm} P{1.5cm} P{1.5cm} }
\toprule
$i$ & \textbf{n} & \textbf{b} & \textbf{SB} & \textbf{SF} & Config-0 & Config-1 & Config-2 \\
\toprule
1  & $(\overline{1}11)$ & $\frac{1}{2}[0\overline{1}1]$ & A2 & 0 & 0.6945 & 0.5536 & 0.4540 \\
2  & $(\overline{1}11)$ & $\frac{1}{2}[101]$ & A3 & 0 & 0.8599 & 0.6544 & 0.5682 \\
3  & $(\overline{1}11)$ & $\frac{1}{2}[110]$ & A6 & 0& 0.8882 & 0.7212 & 0.5563 \\
\rowcolor{lightgray} 4  & $(111)$ & $\frac{1}{2}[0\overline{1}1]$ & B2 & 0.4082 & 0.7711 & {1.0985} & {1.4954} \\
\rowcolor{lightgray} 5  & $(111)$ & $\frac{1}{2}[\overline{1}01]$ & B4 & 0.4082 & 0.6238 & {1.0217} & {1.3904} \\
6  & $(111)$ & $\frac{1}{2}[\overline{1}10]$ & B5 & 0 & 0.8691 & 0.8092 & {1.1442} \\
\rowcolor{lightgray} 7  & $(\overline{1}\overline{1}1)$ & $\frac{1}{2}[011]$ & C1 & 0.4082 & 0.8838 & {1.5671} & {2.4345} \\
\rowcolor{lightgray} 8  & $(\overline{1}\overline{1}1)$ & $\frac{1}{2}[101]$ & C3 & 0.4082 & 0.7518 & {0.9866} & {1.6820} \\
9  & $(\overline{1}\overline{1}1)$ & $\frac{1}{2}[\overline{1}10]$ & C5 & 0 & 0.8541 & 0.5988 & {0.9298} \\
10 & $(1\overline{1}1)$ & $\frac{1}{2}[011]$ & D1 & 0 & 0.7214 & 0.5783 & 0.4366 \\
11 & $(1\overline{1}1)$ & $\frac{1}{2}[\overline{1}01]$ & D4 & 0 & 0.7460 & 0.5217& 0.3388 \\
12 & $(1\overline{1}1)$ & $\frac{1}{2}[110]$ & D6 & 0 & 0.6779 & 0.5183 & 0.4730 \\
\bottomrule 
\end{tabular}
\end{table}

\autoref{tbl:dis_density_110} shows the dislocation densities on individual slip systems during uniaxial deformation along $[110]$ direction, in which only 4 slip systems (B2, B4, C1, C3) are activated with the same Schmid factor.  The remaining 8 slip systems have zero Schmid factor. All slip systems on slip planes A and D are inactive.
\autoref{tbl:dis_density_100} shows that the dislocation densities on the active slip systems increase with strain for both \revision{Config}-1 (strained to 0.07\%) and \revision{Config}-2 (strained to 0.22\%).
On the inactive slip systems B5 and C5, the dislocation densities first decrease (in \revision{Config}-1) and eventually increase (in \revision{Config}-2), due to slip-free multiplication (because their coplanar slip systems B2, B4, and C1, C3 are active).
On the other hand, all slip systems on slip planes A and D are inactive; consequently, slip-free multiplication does not occur for these slip systems and their dislocation densities only decrease with strain.
We conclude that the latent hardening produced by plastic deformation along $[110]$ crystal orientation is due to the highly uneven dislocation densities on different slip planes. 
For example, if a sample first deformed along the $[110]$ direction is unloaded and then reloaded along the $[\bar{1}10]$ direction, then the slip systems A2, A3, D1, D4 will be activated; dislocations on these slip systems will interact strongly with existing ones on B2, B4, C1, C3 through glissile and collinear junctions.
\revision{In our previous work \citep{akhondzadeh2020jmps}, it was reported that the interaction coefficients between slip system pairs that can form glissile and collinear junctions are the highest among all slip system pairs.  This is consistent with the high latent hardening rate for a sample deformed first along the $[110]$ direction and then along the $[\bar{1}10]$ direction.}

\subsection{\revision{Effect of microstructure initialization on yield surface}}
\label{sec:4.2}

\revision{The initial dislocation microstructure is expected to affect the yielding and strain hardening, and hence the yield surface and its evolution. The influence of the initial microstructure can be categorized into three aspects: (1) microstructures with different total dislocation densities, (2) microstructures with nearly the same total dislocation densities but different dislocation distributions on individual slip systems, and (3) microstructures with similar dislocation densities at individual-slip-system level but different dislocation configurations.  These three aspects are discussed one by one in the following.}

\revision{For microstructures with different total dislocation densities, their effects are partially addressed by yield surface calculations of Config-1 and Config-2 from [100] uniaxial loading simulations.  From Config-0 to Config-1 to Config-2, the total dislocation density increases and the yield surface expands isotropically. We note that the increase of dislocation density here (from Config-0 to Config-1 to Config-2) is caused by the dislocation multiplication process during plastic deformation (strain hardening). On the other hand, we can also try to change the dislocation density in the randomly generated initial configuration (Config-0) and examine its effect on the yield surface. We note that the initial dislocation configurations that we generated tend to have a different network structure than dislocation configurations obtained from strain hardening simulations. For example, the initial dislocation configurations tend to have longer dislocation lines. The effect of changing the initial dislocation density on the yield stress in uniaxial loading has been examined in our previous DDD simulations of uniaxial loading \citep{akhondzadeh2023acta}. For $[100]$ loading, increasing the initial dislocation density appears to not influence the initial yield stress significantly but does increase the strain hardening rate. For $[110]$ loading, on the other hand, increasing the initial dislocation density appears to not influence the strain hardening rate (which stays very low) but does increase the initial yield stress. Based on these observations, we expect that if the initial dislocation density (for Config-0) is increased, the expansion of the initial yield surface would depend on the stress orientation.}

\revision{For microstructures with nearly the same total dislocation densities but different distributions on individual slip systems, the corresponding yield surfaces would be quite different. This can be seen by comparing the yield surfaces of [100] Config-1 and [110] Config-1 in \autoref{fig:five_YS}. These two configurations have similar total dislocation densities ($\sim 1.3 \times 10^{12} \mathrm{~m}^{-2}$) but different distributions on individual slip systems (\autoref{tbl:dis_density_100} and \autoref{tbl:dis_density_110}), and their yield surfaces look quite different.}

\begin{table}[h!] 
\centering
\small
\caption{\revision{Dislocation densities (in $10^{11} \mathrm{~m}^{-2}$) for the 12 slip systems in various dislocation configurations (Config-0 and Config-0-comparison) generated using the same approach but different random seeds. Slip system index: $i$, Slip plane: \textbf{n}, Burgers vector: \textbf{b}, Schmid-Boas notation: \textbf{SB}}} \label{tbl:dis_density_comp}
\begin{tabular}{ P{1.0cm} P{1.2cm} P{1.2cm} P{1cm} P{1.5cm} P{3.3cm} }
\toprule
\revision{$i$} & \revision{\textbf{n}} & \revision{\textbf{b}} & \revision{\textbf{SB}} & \revision{Config-0} & \revision{Config-0-comparison} \\
\toprule
\revision{1}  & \revision{$(\overline{1}11)$} & \revision{$\frac{1}{2}[0\overline{1}1]$} & \revision{A2} & \revision{0.6945} & \revision{0.7272} \\
\revision{2}  & \revision{$(\overline{1}11)$} & \revision{$\frac{1}{2}[101]$} & \revision{A3} & \revision{0.8599} & \revision{0.9476} \\
\revision{3}  & \revision{$(\overline{1}11)$} & \revision{$\frac{1}{2}[110]$} & \revision{A6} & \revision{0.8882} & \revision{0.8751} \\
\revision{4}  & \revision{$(111)$} & \revision{$\frac{1}{2}[0\overline{1}1]$} & \revision{B2} & \revision{0.7711} & \revision{0.8997} \\
\revision{5}  & \revision{$(111)$} & \revision{$\frac{1}{2}[\overline{1}01]$} & \revision{B4} & \revision{0.6238} & \revision{0.7555} \\
\revision{6}  & \revision{$(111)$} & \revision{$\frac{1}{2}[\overline{1}10]$} & \revision{B5} & \revision{0.8691} & \revision{0.8194} \\
\revision{7}  & \revision{$(\overline{1}\overline{1}1)$} & \revision{$\frac{1}{2}[011]$} & \revision{C1} & \revision{0.8838} & \revision{0.6176} \\
\revision{8}  & \revision{$(\overline{1}\overline{1}1)$} & \revision{$\frac{1}{2}[101]$} & \revision{C3} & \revision{0.7518} & \revision{0.7071} \\
\revision{9}  & \revision{$(\overline{1}\overline{1}1)$} & \revision{$\frac{1}{2}[\overline{1}10]$} & \revision{C5} & \revision{0.8541} & \revision{0.6563} \\
\revision{10} & \revision{$(1\overline{1}1)$} & \revision{$\frac{1}{2}[011]$} & \revision{D1} & \revision{0.7214} & \revision{0.6507} \\
\revision{11} & \revision{$(1\overline{1}1)$} & \revision{$\frac{1}{2}[\overline{1}01]$} & \revision{D4} & \revision{0.7460} & \revision{0.6787} \\
\revision{12} & \revision{$(1\overline{1}1)$} & \revision{$\frac{1}{2}[110]$} & \revision{D6} & \revision{0.6779} & \revision{0.5800} \\
\bottomrule 
\end{tabular}
\end{table}

\begin{figure}[h!]
    \centering
    \includegraphics[width=1.0\textwidth]{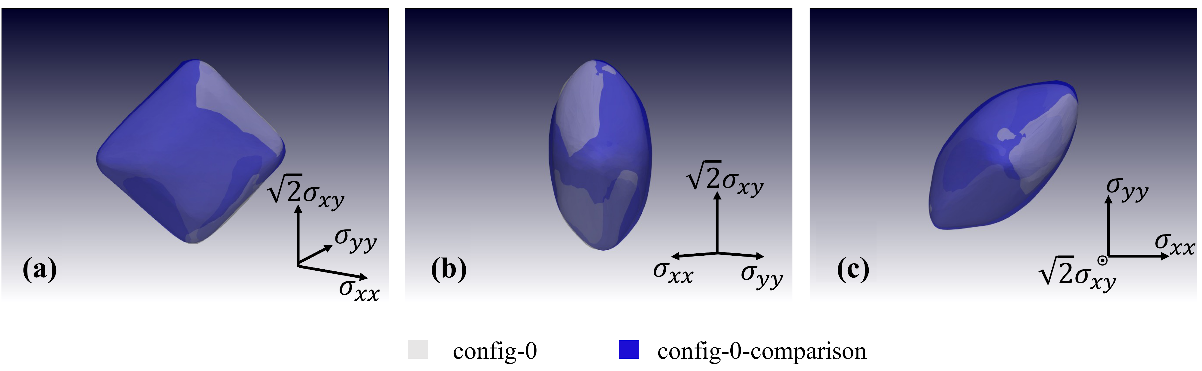}
    \caption{\revision{Yield surfaces of \revision{Config}-0 and \revision{Config}-0-comparison (plotted on top of each other) in three-dimensional plane stress space of ($\sigma_{xx}$, $\sigma_{yy}$, $\sqrt{2}\sigma_{xy}$), in three different viewing angles.}}
    \label{fig:config0_YS_comp}
\end{figure}

\begin{figure}[h!]
    \centering
    \includegraphics[width=0.5\textwidth]{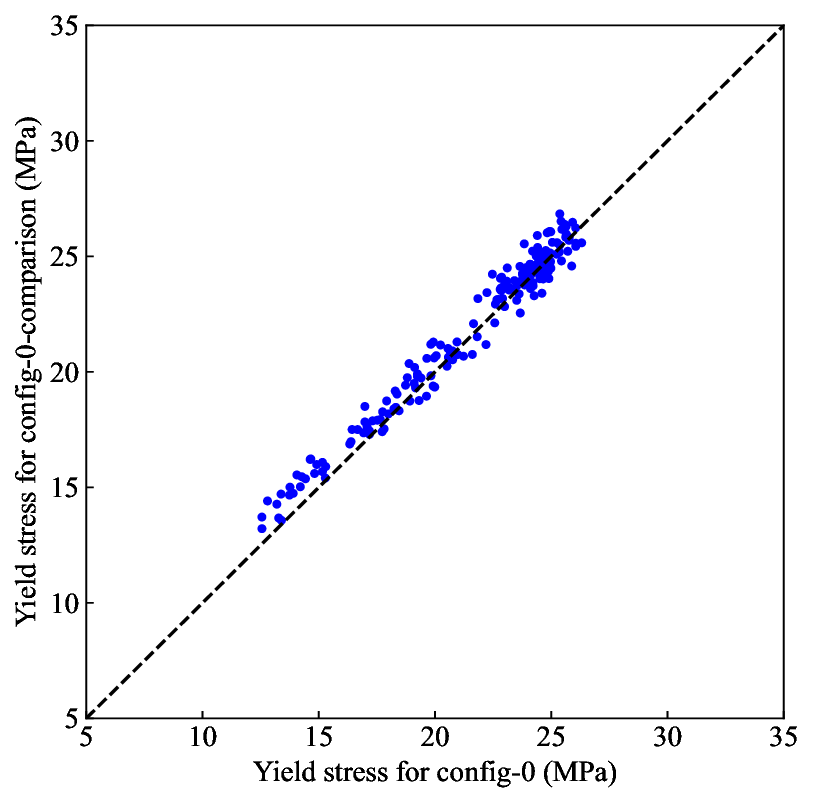}
    \caption{\revision{Yield stress of Config-0-comparison vs that of Config-0 for each stress orientation in the DDD simulations for constructing the yield surface. Each data point corresponds to a specific stress orientation under plane-stress loading.}}
    \label{fig:plot_YS_comp}
\end{figure}

\revision{For microstructures with similar dislocation densities at individual-slip-system level but different dislocation configurations (that are randomly generated), it is expected their yield surfaces are quite similar. Support for this expectation can be found in our earlier DDD simulations of uniaxial loading along the $[100]$ orientation~\citep{sills2018prl}. Three independent dislocation configurations were (randomly) generated.  After initial relaxation, the resulting dislocation densities for all three configurations are around $\rho_0 \approx 0.7 \times 10^{12} \mathrm{~m}^{-2}$. DDD simulations from these three initial configurations predict not only similar yield stress but also consistent stress-strain curves (e.g. strain hardening rates). To explicitly confirm the expectation, we generated another random realization of dislocation configuration (Config-0-comparison). \autoref{tbl:dis_density_comp} shows that the dislocation density on each slip system of this configuration is similar to that of Config-0. We performed DDD simulations to predict the yield surface of Config-0-comparison. \autoref{fig:config0_YS_comp} shows that the predicted yield surface for Config-0-comparison is nearly indistinguishable from that of the Config-0. \autoref{fig:plot_YS_comp} plots the yield stress of Config-0-comparison against the yield stress of Config-0 for each stress orientation. The two agree with each other very well (within statistical error). Therefore, we conclude that the yield surface is insensitive to the (random) initialization of dislocation configurations if the total dislocation density and the density distributions at individual-slip-system level are similar.}

\subsection{\revision{Comparison between DDD-predicted and classical yield surfaces}}
\label{sec:4.3}

%
\revision{It is instructive to compare the DDD-predicted yield surface constructed here with the classical yield surfaces, such as those based on the von Mises and Tresca yield criteria.}
%
\revision{It should be noticed that these two yield surfaces, although widely used to model marcoscopic behaviors of metals, are intended for isotropic materials and hence insensitive to the principal direction of the Cauchy stress. 
Because the DDD predictions in this work are for single crystals, in which plastic deformation occurs by slip on specific crystallographic planes, the corresponding material behavior is inherently anisotropic.}
%
\revision{Therefore, we cannot plot the DDD-predicted yield surface in the space of principal stresses in which von Mises and Tresca yield surfaces are usually plotted.  But we can project the von Mises and Tresca yield surfaces onto the space spanned by the non-zero stress components ($\sigma_{xx}$, $\sigma_{yy}$, $\sqrt{2} \sigma_{xy}$) under plane stress condition to establish a comparison.} 

\begin{figure}[h!]
    \centering
    \includegraphics[width=1.0\textwidth]{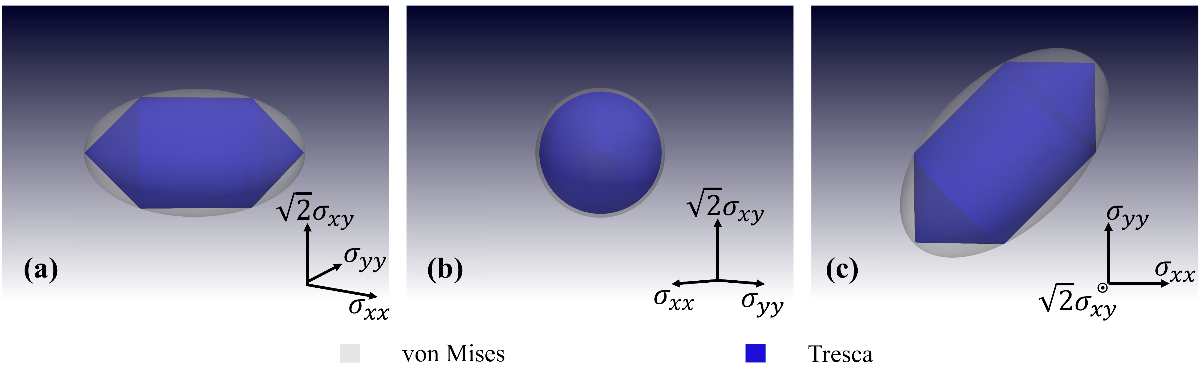}
    \caption{\revision{Yield surfaces of von Mises and Tresca criteria in a three-dimensional plane stress space of ($\sigma_{xx}$, $\sigma_{yy}$, $\sqrt{2}\sigma_{xy}$) in three different viewing angles. The critical yield stress used in both criteria is $\sigma_Y = 16.6$ MPa.}}
    \label{fig:YS_comp}
\end{figure}

\begin{figure}[h!]
    \centering
    \includegraphics[width=0.6\textwidth]{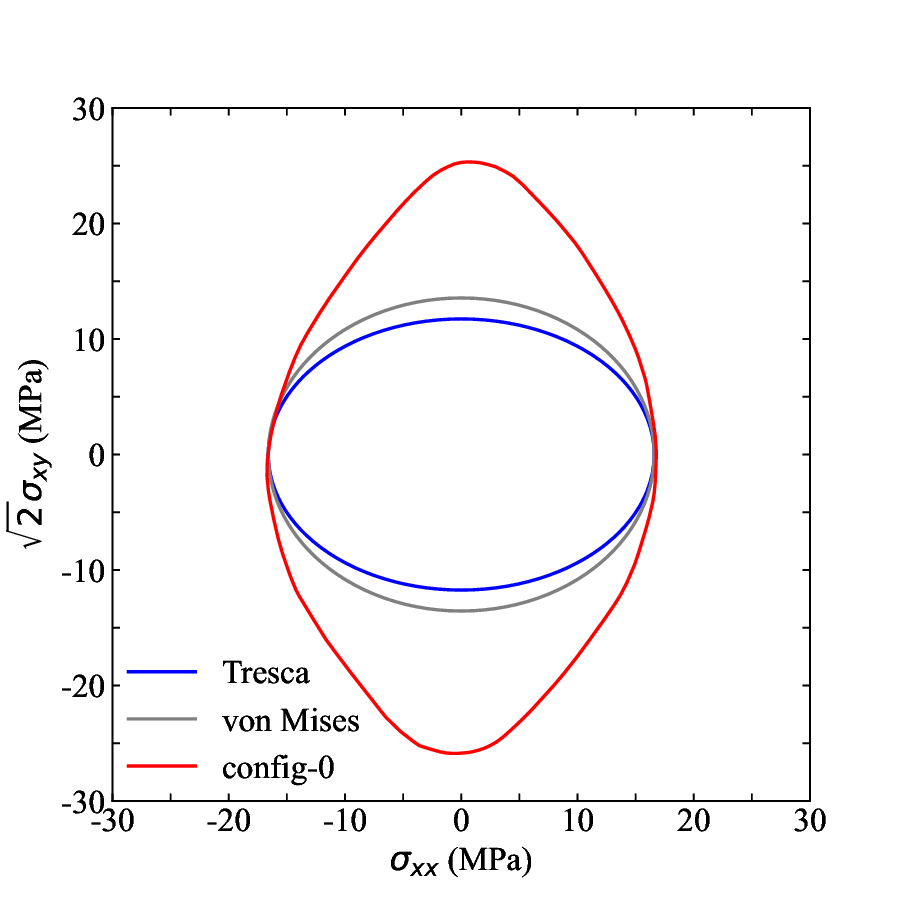}
    \caption{\revision{The cross sections on the $\sigma_{yy} = 0$ plane of the yield surfaces (under plane stress) of von Mises (gray) and Tresca (blue), and DDD prediction of Config-0 (red).
    }}
    \label{fig:2D_YS_comp}
\end{figure}

\revision{\autoref{fig:YS_comp} plots the yield surfaces of von Mises and Tresca in the 3D stress space of plane stress using the yield stress value (for uniaxial loading) of $\sigma_{\rm Y} = 16.6$~MPa. Visual inspection clearly indicates that both these yield surfaces have rather different shapes compared with the yield surfaces from DDD simulations for single crystals. 
For example, both the von Mises and the Tresca yield surfaces resemble a circle when viewed in the angle of \autoref{fig:YS_comp}(b), but the single crystal yield surface appears very anisotropic in the same viewing angle shown in \autoref{fig:five_YS}(b).}
\revision{The difference is also clearly seen in \autoref{fig:2D_YS_comp}, which plots the cross-section of the yield surface on the plane of $\sigma_{yy} = 0$.
This is the space of tension and shear that can be probed by combined tensile and torsional loading of thin-walled tubes \citep{taylor1931ptrsa}.}
\revision{The difference between DDD constructed and classical yield surfaces is caused by the fact that, in FCC metals such as Cu, yielding occurs by slipping on $\{111\}$-type planes, hence a much higher shear stress is required if the maximum shear stress is on the $\{100\}$-type planes (for single crystals). On the other hand, in von Mises or Tresca theories that are developed for isotropic materials (e.g. polycrystals), yielding occurs independent of the plane orientation of the shear stresses.}

\subsection{Next steps in DDD simulations}
\label{sec:4.4}

In this work, we have demonstrated that DDD simulations can be used to construct the yield surface of single-crystal Cu under plane-stress conditions and how the yield surface evolves under uniaxial plastic deformation along $[100]$ and $[110]$ crystal orientations.
Due to computational limits on the simulation time scale, the yield surface calculations are calculated under a stress rate of $10^{13} \, {\rm Pa}\cdot{\rm s}^{-1}$, which is equivalent to the strain rate of $10^3 \, {\rm s}^{-1}$ for the uniaxial deformation.
However, what we usually have in mind when discussing yield surface or strain-hardening behaviors as well as most of the existing experimental data pertain to much lower, quasi-static strain rates, ranging from $10^{-4} \, {\rm s}^{-1}$ to $10^{-1} \, {\rm s}^{-1}$~\citep{meyers1994dynamic,meyers2008mechanical}.
Since strain-rate effect on the plastic deformation exists in most of the materials~\citep{meyers1994dynamic}, including polymers \citep{farrokh2010ejma}, metals \citep{butcher1966jap,meyers2012shock} and composites \citep{sierakowski1997amr}, it is expected that the yield surface of single crystal copper is strain-rate dependent.
At present, lowering the strain rate of DDD simulations to the level of, say $10^{-3} \, {\rm s}^{-1}$, by brute force appears out of reach.
Fortunately, existing experimental data \citep{edington1969pm} suggests that the flow stress of single-crystal Cu appears to be strain-rate independent in a wide range of strain-rates from $10^{-4} \, {\rm s}^{-1}$ to $10^{1} \, {\rm s}^{-1}$.
DDD simulations of single-crystal Cu at the strain rate of $10^{2} \, {\rm s}^{-1}$ have been reported earlier \citep{sills2018prl}.  Hence, DDD simulations at the strain rate of $10^{1} \, {\rm s}^{-1}$ appear feasible in the near future, with a bigger commitment of computational resources.
A challenge that one may need to face is that the size of the DDD simulation cell is likely to increase in order to reach convergence when the strain rate is lowered, which would further increase the computational cost.

In addition, more DDD simulations would be needed if we wish to go beyond the confines of plane-stress loading conditions and construct the yield surface in the \revision{$n$-dimensional stress space with $n > 3$. The greatest challenge in a brute force approach is the large number of DDD simulations along different stress orientations required so that the geometry of the yield surface in the stress space can be described well. The number $N$ of independent DDD simulations required is roughly the number of points needed to adequately sample a $(n-1)$-dimensional unit sphere in $n$-dimensional space.}
%
%
%
%
\revision{The necessary number of sampling points can be estimated by $N \approx S_{n-1} / \lambda^{n-1}$, where $S_{n-1} = {2 \pi^{n/2}} / {\Gamma(n/2)}$ is the area of a unit $(n-1)$-dimensional sphere~\citep{smith1989mm}, and $\lambda$ is the distance between nearest sampling points on the unit sphere.  For example, we may choose $\lambda \approx 0.3$, then for $n = 3$ (e.g. plane stress), $N \approx 140$, which is consistent with the number of DDD simulations used in this work to construct each yield surface. }
\revision{However, if we were to sample the full 6D stress space ($n = 6$) by brute force, it would require $N \approx 12760$ sampling points, which amounts to two orders of magnitude more DDD simulations than those performed in this work.  With the advances in computing power, such calculations may be feasible in the near future; but they are beyond the scope of this work.}

\revision{In addition,} it appears necessary to repeat DDD simulations under the same stress conditions for different initial dislocation configurations to improve statistics. \revision{Therefore, more efficient sampling of the stress space is necessary for constructing the yield surface in full 6D from DDD simulations.}

\subsection{Next steps in geometric prior method}
\label{sec:4.5}

We have demonstrated that the geometric prior method can be used to construct the yield surface as a manifold in stress space based on data obtained from DDD simulations.
So far, the manifolds corresponding to each dislocation configuration (e.g. [100] \revision{Config}-1, [100] \revision{Config}-2, etc.) are constructed independently of each other, even though they are related to each other through plastic deformation starting from a common configuration (\revision{Config}-0).
As a natural next step, it would be of interest for the geometric prior method to learn not only the yield surfaces separately but its evolution with plastic deformation.
Such a model is ultimately needed in a  constitutive model of crystal plasticity at the continuum scale.

Intuitively, a starting point for modeling the evolution of the yield surface is to generate enough snapshots of the yield surface at different plastic deformation levels. However, the number of yield surfaces that can be obtained from DDD simulations is limited due to the high computational cost. 
Hence an interpolation scheme is needed between sparse data points along the strain axis.
Based on differential manifold theory \citep{kreyszig2013differential}, the interpolation between two yield surfaces should be rigorously defined by a transformation mapping from one manifold to another one. To derive an interpolation policy from a manifold transformation, it is necessary that the two underlined surfaces are isomorphic, i.e., it is required to establish the one-to-one correspondence between two points on these two surfaces. 
This appears to be a valid assumption at the strain level accessible to DDD simulations in the near future.

For the geometric prior method used in this study, we use a few patches to describe the local features of a yield surface. The patches can overlap and then provide a complete description of the entire yield surface. 
To ensure an accurate description of yield surface in this work, we have tried to create altas consisting of 
different numbers of patches to construct yield surfaces of different complexity. The total number of patches varies from 30 to 37. 
Note that, if two yield surfaces are represented by two altas consisting of different numbers of patches, it is not trivial to generate an injective-mapping between them to interpolate a hardening law. 
Therefore, we set the yield surfaces from two configurations along the same deformation trajectory to be represented by the altas consisting of the same number of patches.
Along this line, one feasible approach is to simplify the calculation is to assume that the 
motions of individual patches in the stress space (and hence the hardening/softening behaviors) can be characterized by a combination of kinematics, such as translation, scaling, and rotation.
This would allow us to construct a plasticity evolution law by establishing relationships between the patch motion parameters and the cumulative plastic strain \citep{xiao2022cmame}.
Meanwhile, one must be cautious that the trade-off of the flexibility afforded by the geometric prior approach is the additional overhead training required to maintain consistency between patches in the overlapping regions in this new scheme \citep{xiao2022cmame}.

To extend the yield surface to the full stress space (6D), one promising approach would be turning back to the implicit yield function representation but fitting it with the neural kernel method \citep{xiong2023cmame} such that the yield surface is constituted by a smooth basis extracted from kernel regression. This generally produces a smoothed yield surface with good generalizability, i.e., the yield surface will not deform to complex shapes locally due to insufficient training data, but some accuracy will be lost at locations with real geometric complexities. Essentially, in more-than-three-dimensional \revision{non-plane-stress} spaces, hyper-surface reconstruction is still a challenging task, and the trade-off between local accuracy and global generalizability is more difficult to make than in \revision{the 3D stress space of plane stress}.

\section{Conclusions}
\label{sec:conclude}

In this work, we demonstrate a framework to construct the yield surface of single crystals using DDD simulations and the geometric prior method.
DDD simulations under constant stress-rate conditions have been performed to identify the yielding conditions and the data are used by the geometric prior method to construct a cross-section of the yield surface as a manifold in the 3-dimensional sub-space of plane stress.
An iterative workflow is adopted, where the geometric prior method identifies regions of interest (based on local curvature) where further sampling of the stress conditions by DDD simulations is performed. 

We found that the yield surface evolves differently by plastic deformation along the $[100]$ and $[110]$ loading directions, respectively.  Isotropic hardening is observed for $[100]$ deformation, in which the yield surface expands with strain by nearly the same ratio in all stress orientations.
This is traced to dislocation multiplication on all four slip planes during $[100]$ deformation.
In contrast, latent hardening is observed for $[110]$ deformation, in which the yield surface does not expand much at all in the vicinity of the stress orientations corresponding to $[110]$ tension but expands significantly in other orientations.
This is traced to dislocation multiplication on only two out of the four slip planes during $[110]$ deformation.

\revision{The code and data necessary to reproduce all the results in this paper are available on Github~\citep{GDLYieldSurf}. For each yield surface constructed here, a yield criterion is furnished by a Python program which determines whether a given plane stress condition is inside, exactly on, or outside the yield surface.}

\section{Acknowledgements}
This work was supported by the National Science Foundation under Award Number DMREF 2118522 (W.J. and W.C.). W.C.S. and X.M. are supported by the UPS Foundation Visiting Professorship from Stanford University, with additional support from the National Science Foundation under grant contracts CMMI-1846875 and the Dynamic Materials and Interactions Program from the Air Force Office of Scientific Research under grant contracts FA9550-21-1-0391 and FA9550-21-1-0027, with additional funding from the Department of Energy DE-NA0003962.

\setcounter{figure}{0}
\setcounter{table}{0}

\appendix
\section{Constant stress rate vs constant strain rate loading}
\label{sec:appendix-A}
\begin{figure}[htb]
    \centering
    \includegraphics[width=0.5\textwidth]{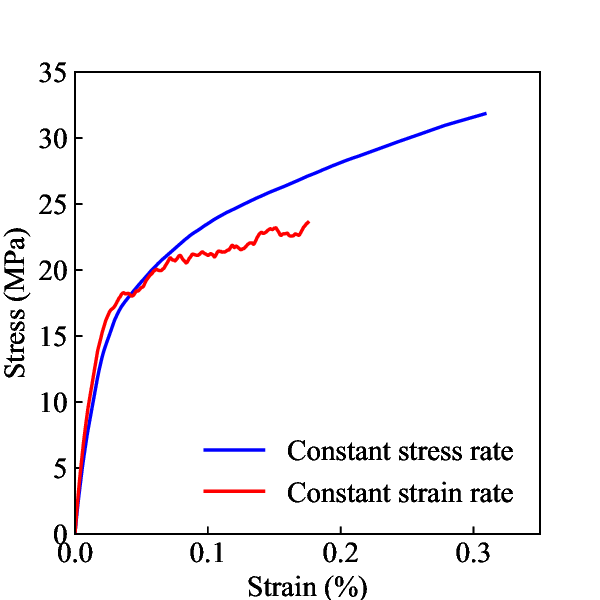}
    \caption{The stress-strain curves for uniaxial tensile loadings with constant stress rate of $10^{13} \, {\rm Pa}\cdot{\rm s}^{-1}$ and constant strain rate of $10^3 \, {\rm s}^{-1}$ along the $[100]$ crystal orientation, respectively.}
    \label{fig:stress_strain_curves_comparison}
\end{figure}

The comparison between the stress-strain curves for uniaxial tensile loading with constant stress rate of $10^{13} \, {\rm Pa}\cdot{\rm s}^{-1}$ and constant strain rate of $10^3 \, {\rm s}^{-1}$ along the $[100]$ orientation is shown in \autoref{fig:stress_strain_curves_comparison}.

\revision{The stress-strain curves of a DDD simulation under constant strain rate loading are usually not monotonic, but show up-and-downs due to statistical fluctuations of dislocation activities.}
\revision{On the other hand, in the constant stress rate simulation, the stress is stipulated to increase monotonically with time.  As a result, the stress-strain curve looks much smoother.  Here, the constant stress rate simulation is only a convenient way for us to obtain the yield stress from DDD simulations.  Despite the differences in the stress-strain curves between constant strain rate and constant stress rate conditions in the post-yield region, they are close to each other in their predictions of the yield stress, which marks the transition from the initial elastic regime to the post-yield plastic regime.}

\section{DDD simulation parameters}
\label{sec:appendix-B}
The parameters of our DDD simulations are summarized in \autoref{tbl:simParam}.  

\begin{table}[h!] 
\centering
\small
\caption{Summary of DDD simulation parameters for single crystal copper.} \label{tbl:simParam}
\arrayrulecolor{black}
\begin{tabular}{ P{5.5cm}  P{1.6cm}  P{3.9cm}  }
\toprule
Property & Parameter & Value  \\
\toprule
Shear modulus & $\mu$ & 54.6 GPa  \\
Poission's ratio & $\nu$ & 0.324   \\
Burgers vector magnitude & $b$ & 0.255 nm   \\
Drag coefficient & $B$ & 15.6 $\mathrm{\mu~Pa} \cdot \mathrm{s}$   \\
Core radius & a & 6 $b$   \\
Relative tolerance &  $f_{\rm tol}$   &   0.1 $b$   \\
Absolute tolerance &  $r_{\rm tol}$   &   10 $b$   \\
Threshold tolerance & $r_{\rm th}$    &   1 $b$   \\
Subcycling group 1 radius & $r_{g1}$  &   0     \\
Subcycling group 2 radius & $r_{g2}$  &   100 $b$   \\
Subcycling group 3 radius & $r_{g3}$  &   600 $b$   \\
Subcycling group 4 radius & $r_{g4}$  &   1600 $b$  \\
Max segment length &  $l_{\rm max}$   &   2000 $b$   \\
Min segment length &  $l_{\rm min}$   &   $\sqrt{\frac{4}{\sqrt{3}} A_{\min }}$   \\
Max area between segments &  $A_{\rm max}$  &   $\frac{1}{2}\left(4 A_{\min }+\frac{\sqrt{3}}{4} l_{\max }^2\right)$   \\
Min area between segments &  $A_{\rm min}$  &   $\min \left(2 r_{\mathrm{tol}} l_{\max }, \frac{\sqrt{3}}{4} l_{\min }^2\right)$   \\
Collision radius &   $r_{\rm col}$  &    10 $b$  \\
Simulation cell size &   $L$  &  15 $\mu$m    \\
Strain rate for uniaxial loading &    $\dot{\varepsilon}$      &   $10^3 \, {\rm s}^{-1}$   \\
Stress rate for plane stress loading &   $\dot{\left\| \sigma \right\|}$     &   $10^{13} \, {\rm Pa}\cdot{\rm s}^{-1}$   \\
\bottomrule 
\end{tabular}
\end{table}


\bibliographystyle{model2-names}

\end{document}